%% file: main_arxiv.tex
\documentclass[3p]{article}

\usepackage[utf8]{inputenc}

\newcommand*{\qed}{\hbox{}\hfill$\Box$}
\newcommand*{\qedhere}{\qed}
\newenvironment{proof}{{\noindent \emph{Proof.}}}{}

\def\TypeOfVersion{arxiv}
\input{macros_preliminary_macros.tex}

\usepackage{amsmath}
\usepackage{amssymb}
\usepackage{booktabs} 
\usepackage{comment}
\usepackage{xifthen}
\usepackage{textcase}
\usepackage{fullpage}
\usepackage{enumitem}
\usepackage{algorithmicx,algorithm}
\usepackage[noend]{algpseudocode}
\usepackage{tikz} \usetikzlibrary{calc,snakes}
\usepackage{etoolbox}
\usepackage{subfigure,color,soul}
\usepackage{multirow}

\usepackage{hyperref}
\hypersetup{pdfauthor={Ricardo C. Corrêa and Pablo M. S. Farias},
            citecolor=blue,        		
            colorlinks=true,       		
            filecolor=blue,      		
            linkcolor=blue,          	
            urlcolor=blue,
           }

\input{macros.tex}
\newcommand{\CALL}[2]{\ensuremath{\textsc{#1}\ifthenelse{\isempty{#2}{}}{}{(#2)}}}

\title{Linear Time Computation of the Maximal Linear and Circular Sums of
Multiple Independent Insertions into a Sequence}

\author{{\sc Ricardo C. Corrêa} \\ {\small Universidade Federal Rural do Rio de
Janeiro} \\ {\small Departamento de Ciência da Computação} \\
{\small Av. Governador Roberto Silveira S/N} \\
{\small 26020-740 Nova Iguaçu, RJ, Brazil} \\ {\small \tt correa@ufrrj.br} \and
{\sc Pablo M. S. Farias} \\ {\small Universidade Federal do Ceará} \\ {\small
Departamento de Computação} \\ {\small Campus do Pici, Bloco 910} \\
{\small 60440-554 Fortaleza, CE, Brazil} \\
{\small \tt pmsf@lia.ufc.br}}

\date{}

\begin{document}

\maketitle

\begin{abstract}
The \emph{maximal sum} of a sequence $A$ of $n$ real numbers
  is the greatest sum of all elements of any linearly contiguous
  and possibly empty subsequence of $A$. It can be computed in \Oof{n} time by
  means of Kadane's algorithm.
Letting \seqins[x]{A}{p} denote the sequence which results
  from inserting a real number $x$ just after element \seqindex{A}{p-1}, we show how the maximal sum of \seqins[x]{A}{p} can be computed in \Oof{1} \emph{worst-case} time for any given $x$ and $p$,
  provided that an \Oof{n} time preprocessing step has already been executed on $A$.
In particular, this implies that,
  given $m$ pairs $(x_0, p_0), \ldots, (x_{m-1}, p_{m-1})$,
  we can compute the maximal sums of sequences
  $\seqins[x_0]{A}{p_0}, \ldots, \seqins[x_{m-1}]{A}{ p_{m-1} }$
  optimally in \Oof{n+m} time,
  improving on the straightforward and suboptimal strategy of applying Kadane's
  algorithm to each sequence \seqins[x_i]{A}{p_i}, which takes a total of \Thetaof{nm} time.
We also show that the same time bound is attainable when circular subsequences
of \seqins[x]{A}{p} are taken into account.
Our algorithms are easy to implement in practice,
  and they were motivated by a buffer minimization problem on wireless mesh networks.
\end{abstract}

\input{intro}
\input{linear}
\input{circular}
\input{SecUnitConclusion.tex}

\section*{Acknowledgment}

This work is partially supported by FUNCAP/INRIA (Ceará State, Brazil/France) and CNPq (Brazil) research projects.
We are very grateful for the invaluable remarks by the anonymous referee, in
special for bringing the algorithms due to Lin et al.~\cite{BibLinJiangChao2002}
and Mu~\cite{BibMu2008}, as well as Jeuring's algorithm~\eqref{eq:noncircposx},
to our attention.

\bibliographystyle{unsrt}
\bibliography{./bibliography}

\end{document}

%% file: macros_preliminary_macros.tex
\newcommand{\InsertSubmissionOrSelfInput}
  [2] 
  {\ifthenelse{\equal{\TypeOfVersion}{submission}}{#1}{#2}}

%% file: macros.tex
\input{macros_algorithms.tex}
\input{macros_arrayDrawing.tex}
\input{macros_crossref.tex}

\input{macros_math.tex}
\input{macros_over-underbrace_in_tikz.tex}
\input{macros_text.tex}

\newcommand{\calculate} 
  [1]
  {\pgfmathparse{#1}\pgfmathprintnumber{\pgfmathresult}}

\newcommand{\EdefAsResultOf} 
  [2] 
  {\pgfmathparse{#2}\edef#1{\pgfmathresult}}

\def\shouldinsertfigures{yes}
\newcommand{\insertfigure}
  [1] 
  {\ifthenelse{\equal{\shouldinsertfigures}{yes}}{\input{figures/#1}}{\mbox{\reminder{---OMMITED FIGURE!---}}}}

\makeatletter
\newtoks\@tabtoks
\newcommand\addtabtoks[1]{\global\@tabtoks\expandafter{\the\@tabtoks#1}}
\newcommand\eaddtabtoks[1]{\edef\mytmp{#1}\expandafter\addtabtoks\expandafter{\mytmp}}
\newcommand*\resettabtoks{\global\@tabtoks{}}
\newcommand*\printtabtoks{\the\@tabtoks}
\makeatother

%% file: macros_algorithms.tex
\newcommand{\algorithmname} 
  [2] 
  [-]
  {\ifthenelse{\equal{#1}{-}}{\textsc{#2}}{\mbox{\textsc{#2}}}}

\newcommand{\defaultValueOfAssign}{newline}
\newcommand{\assign} 
  [3] 
  [\defaultValueOfAssign]
  {#2 \ensuremath{\leftarrow} #3\ifthenelse{\equal{#1}{;}}
                                           {\semiclsameline}
                                           {\ifthenelse{\equal{#1}{}}{}{ \;}}}%

\newcommand{\assignWithOperation} 
  [4][\defaultValueOfAssign] 
  {\assign[#1]{\ensuremath{#2}}{\ensuremath{#2 #3 #4}}}

\newcommand{\callObjectMethod}
  [2] 
  {\objfield{#1}{#2}}

\newcommand{\defineMethod}
  [4][notinline] 
  {\KwSty{method}
   \objWithType{#2}{\ifthenelse{\equal{#3}{none}}
                               {\KwSty{no\_return}}
                               {#3}}
   \ifthenelse{\equal{#1}{inline}}
              {\ \,\KwSty{begin} #4 \KwSty{end}}
              {\; \Begin{#4}}}

\newcommand{\fieldname} 
  [1] 
  {\ensuremath{\text{\textnormal{#1}}}}

\newcommand{\objfield} 
  [2] 
  {#1\mbox{\textnormal{\textbf{.}}}\fieldname{#2}}

\newcommand{\functioncall}[3][-] 
  {\algorithmname[#1]{#2}\ensuremath{(#3)}}

\newcommand{\objWithType}
  [2] 
  {\mbox{#1\::\:#2}}

\newcommand{\recordStructure}
  [1] 
  {\ensuremath{\{}\,#1\,\ensuremath{\}}}

\newcommand{\semiclsameline}{\,; }

\newcommand{\typesetTypeName}
  [1] 
  {\ensuremath{\text{\textnormal{#1}}}}

\newcommand{\varname} 
  [1] 
  {\ensuremath{\mathit{#1}}} 

%% file: macros_arrayDrawing.tex
\edef\stdFillingColor{gray!20}

\edef\elemXSide{0.76} 
\edef\elemYSide{0.67}
\edef\arrayXPos{0}
\edef\arrayYPos{0}
\edef\arrayNumElem{8} 
\edef\otherArrayDrawingOpt{}

\newcommand{\drawEmptyArray}
  {\draw[\otherArrayDrawingOpt] (\arrayXPos,\arrayYPos) rectangle (\arrayXPos + \arrayNumElem*\elemXSide,\arrayYPos + \elemYSide);}

\edef\ExtraSpaceBtwArrayAndItsLabel{0}
\newcommand{\putLabelOfArray}
  [1]
  { \node[left] at (\arrayXPos - \ExtraSpaceBtwArrayAndItsLabel,\arrayYPos + \elemYSide/2) {#1}; }

\pgfmathparse{0}
\edef\spaceBtwIndexArray{\pgfmathresult}
\newcommand{\putIndexInArrayAtX}
  [2] 
  {\ifthenelse {\equal{#1}{}} {} {\node[above] at (#2, \arrayYPos + \elemYSide + \spaceBtwIndexArray) {#1}; } }

\newcommand{\putIndexBeforeNode}
  [2] 
  {\pgfmathparse{\arrayXPos + #1*\elemXSide}
   \edef\Xmiddle{\pgfmathresult}
   \putIndexInArrayAtX{#2}{\Xmiddle} }

\newcommand{\putLabelInArrayAtX}
  [2] 
  {\ifthenelse {\equal{#1}{}} {} {\node at (#2, \arrayYPos + \elemYSide/2) {#1}; } }

\newcommand{\putNodeInfo}
  [3] 
  {\pgfmathparse{\arrayXPos + (#1 + 0.5)*\elemXSide}
   \edef\Xmiddle{\pgfmathresult}
   \putLabelInArrayAtX{#2}{\Xmiddle}
   \putIndexInArrayAtX{#3}{\Xmiddle} }

\edef\LittleUnderbraceSizeDecrease{0.03}
\edef\ExtraSpaceBtwArrayAndUnderbrace{0}
\newcommand{\underbraceNodes} 
  [3] 
  {\UnderbraceTikZPoints[\ExtraSpaceBtwArrayAndUnderbrace]
                        {$(\arrayXPos,\arrayYPos)+(#1*\elemXSide + \LittleUnderbraceSizeDecrease,0)$}
                        {$(\arrayXPos,\arrayYPos)+(#2*\elemXSide + \elemXSide - \LittleUnderbraceSizeDecrease,0)$}
                        {#3}}

\edef\drawOfSubseq{black}
\edef\fillOfSubseq{\stdFillingColor}
\edef\otherSubseqDrawingOpt{}
\newcommand{\drawSubseq}
  [4] 
  {\draw[draw=\drawOfSubseq,fill=\fillOfSubseq,\otherSubseqDrawingOpt]
        (\arrayXPos + #1*\elemXSide,\arrayYPos) rectangle
        (\arrayXPos + #2*\elemXSide + \elemXSide,\arrayYPos + \elemYSide);
   \pgfmathparse{\arrayXPos + #1*\elemXSide + (#2 - #1 + 1)*\elemXSide/2}
   \edef\Xmiddle{\pgfmathresult}
   \putLabelInArrayAtX{#3}{\Xmiddle}
   \putIndexInArrayAtX{#4}{\Xmiddle} }

\newcommand{\drawSubseqNoFill}
  [4] 
  {\draw[draw=\drawOfSubseq,\otherSubseqDrawingOpt]
        (\arrayXPos + #1*\elemXSide,\arrayYPos) rectangle
        (\arrayXPos + #2*\elemXSide + \elemXSide,\arrayYPos + \elemYSide);
   \pgfmathparse{\arrayXPos + #1*\elemXSide + (#2 - #1 + 1)*\elemXSide/2}
   \edef\Xmiddle{\pgfmathresult}
   \putLabelInArrayAtX{#3}{\Xmiddle}
   \putIndexInArrayAtX{#4}{\Xmiddle} }

%% file: macros_crossref.tex
\newcommand{\citealg}[2][short] 
  {\ifthenelse{\equal{#1}{long}}{Algorithm}{Alg.}~\ref{#2}}

\newcommand{\citeappendix}[1] 
  {Appendix~\ref{#1}}

\newcommand{\citeequation}[1] 
  {\eqref{#1}}

\newcommand{\citefigure}[1] 
  {Fig.~\ref{#1}}

\newcommand{\citelemma}[1] 
  {Lemma~\ref{#1}}

\newcommand{\citenote}[1] 
  {Note~\ref{#1}}

\newcommand{\citeobservation}[1] 
  {Observation~\ref{#1}}

\newcommand{\citeremark}[1] 
  {Remark~\ref{#1}}

\newcommand{\citesection}[1] 
  {Section~\ref{#1}}

\newcommand{\citesubsection}[1] 
  {Subsection~\ref{#1}}

\newcommand{\citetable}[1] 
  {Table~\ref{#1}}

\newcommand{\citetheorem}[1] 
  {Theorem~\ref{#1}}

\newcommand{\AlgorithmMaxQueue}{\citealg{AlgorithmMaxQueue}}

\newcommand{\SecUnitConclusion}{\citesection{SecUnitConclusion}}

\newcommand{\SecUnitGeneralApproach}{\citesection{SecUnitGeneralApproach}}

\newcommand{\tableNoncirc}{\citetable{tableNoncirc}}

%% file: macros_math.tex
\DeclareMathOperator*{\argmax}{arg\,max}

\newcommand{\arrayindex}
  [2] 
  {\ensuremath{#1[#2]}}

\newcommand{\arrayname}[1]{\ensuremath{\mathit{#1}}}

\def\definenewarray#1#2{
  \expandafter\def\csname array#1\endcsname{\arrayname{#2}}%
  \expandafter\def\csname array#1of\endcsname##1{%
    \arrayindex{\csname array#1\endcsname}
               {##1}}}

\newcounter{casocont}
\newcommand{\caso} 
  [3] 
  [nada]
  {\ifthenelse{\value{casocont}=0}
              {}
              {\\}%
   {#2},\hspace*{1ex}%
   \ifthenelse{\value{chavedecasosalin} = 0}
              {}
              {&}%
   \ifthenelse{\equal{#1}{nada}}
              {\emmodopar{#3}}
              {\emmodopar[#1]{#3}}%
   \addtocounter{casocont}{1}}

\newcounter{chavedecasosalin}
  [1]
  {\left\{\ifthenelse{\equal{#1}{0}}
                     {\begin{array}{@{}l}\setcounter{chavedecasosalin}{0}}
                     {\begin{array}{@{}l@{}l}\setcounter{chavedecasosalin}{1}}%
   \setcounter{casocont}{0}}
  {\end{array}\right.}

\newlength{\emmodoparlargura}
\newcommand{\emmodopar} 
  [2] 
  [\emmodoparlargura]
  {\settowidth{\emmodoparlargura}{{#2}}\begin{minipage}[t]{#1}{#2}\end{minipage}}

\newcommand{\bigO}{\ensuremath{O}}
\newcommand{\littleo}{\ensuremath{o}}
\newcommand{\bigTheta}{\ensuremath{\Theta}}
\newcommand{\littleomega}{\ensuremath{\omega}}

\newcommand{\omegaof} 
  [1] 
  {\unaryfunctionof{\littleomega}{#1}}

\newcommand{\oof} 
  [1] 
  {\unaryfunctionof{\littleo}{#1}}

\newcommand{\Oof} 
  [1] 
  {\unaryfunctionof{\bigO}{#1}}

\newcommand{\Thetaof} 
  [1] 
  {\unaryfunctionof{\bigTheta}{#1}}

\newcommand{\SquareBracketize}
  [2][1] 
  {\delimit{#1}{[}{]}{#2}}

\newcommand{\delimitersize}[1]
  {\ifthenelse{\equal{#1}{1}}
              {}
   {\ifthenelse{\equal{#1}{2}}
               {\big}
    {\ifthenelse{\equal{#1}{3}}
                {\Big}
     {\ifthenelse{\equal{#1}{4}}
                 {\bigg}
      {\ifthenelse{\equal{#1}{5}}
                  {\Bigg}
                  {WRONG-DELIMITER-SIZE}}}}}}

\newcommand{\delimit}
  [4] 
  {\ensuremath{\ifthenelse{\equal{#1}{*}}{\left}{\delimitersize{#1}}#2
               #4
               \ifthenelse{\equal{#1}{*}}{\right}{\delimitersize{#1}}#3}}

\newcommand{\parenthesize}
  [2][1] 
  {\delimit{#1}{(}{)}{#2}}

\newcommand{\binfunctionof}
  [3] 
  {\naryfunctionof{#1}{#2,#3}}

\newcommand{\mathfunctionname}
  [1] 
  {\ensuremath{\mathit{#1}}}

\newcommand{\naryfunctionof} 
  [2] 
  {\ensuremath{#1(#2)}}

\newcommand{\ternaryfunctionof}
  [4] 
  {\naryfunctionof{#1}{#2,#3,#4}}

\newcommand{\unaryfunctionof}
  [2] 
  {\naryfunctionof{#1}{#2}}

\newcommand{\begofinterv}
  [1] 
  {\ensuremath{\alpha_{#1}}}

\newcommand{\Endofinterv}
  [1] 
  {\ensuremath{\alpha_{#1+1}}}

\newcommand{\interv}
  [1] 
  {\ensuremath{I_{#1}}}

\newcommand{\barinterv}
  [1] 
  {\ensuremath{\bar I_{#1}}}

\newcommand{\intervpartition}
  {\seq{ \begofinterv{0}, \begofinterv{1}, \ldots, \begofinterv{\numofinterv-1} }}

\newcommand{\numofinterv}
  {\ensuremath{\ell}}

\newcommand{\plusmodinterv}[2]
  {\ensuremath{#1 \plusmod{ \numofinterv } #2}}

\newcommand{\column}
  [2] 
  {\binfunctionof{\mathit{col}}{#1}{#2}}

\newcommand{\matrixelem}
  [3] 
  {\ensuremath{#1[#2,#3]}}

\newcommand{\matrixrow}
  [2] 
  {\ensuremath{#1_{#2}}}

\newcommand{\row}
  [2] 
  {\binfunctionof{\mathit{row}}{#1}{#2}}

\newcommand{\costofmatrix}
  [1] 
  {\unaryfunctionof{\mathit{score}}{#1}}

\newlength{\colocarabaixoaux}
\newlength{\colocarabaixoauxdois}
\newcommand{\colocarabaixo}
  [3] 
  [0pt]
  {
   \ensuremath{\text{#2}}
   \settoheight{\colocarabaixoaux}{#3}%
   \settodepth{\colocarabaixoauxdois}{#2}%
   \addtolength{\colocarabaixoaux}{\colocarabaixoauxdois}%
   \addtolength{\colocarabaixoaux}{#1}%
   \raisebox{-\colocarabaixoaux}{\makebox[0pt][r]{%
                      \settowidth{\colocarabaixoaux}{#3}%
                      \makebox[0pt][r]{#3\hspace*{-.5\colocarabaixoaux}}%
                      \settowidth{\colocarabaixoaux}{#2}%
                      \hspace*{.5\colocarabaixoaux}%
                      }}}

\newcommand{\colocaracima}
  [3] 
  [0pt]
  {
   \ensuremath{\text{#2}}
   \settodepth{\colocarabaixoaux}{#3}%
   \settoheight{\colocarabaixoauxdois}{#2}%
   \addtolength{\colocarabaixoaux}{\colocarabaixoauxdois}%
   \addtolength{\colocarabaixoaux}{#1}%
   \raisebox{\colocarabaixoaux}{\makebox[0pt][r]{%
                      \settowidth{\colocarabaixoaux}{#3}%
                      \makebox[0pt][r]{#3\hspace*{-.5\colocarabaixoaux}}%
                      \settowidth{\colocarabaixoaux}{#2}%
                      \hspace*{.5\colocarabaixoaux}%
                      }}}

\newcommand{\operacaomodulo} 
  [4] 
  [\scriptsize]
  {\ensuremath{\colocarabaixo[#4]{$#2$}{{#1$#3$}}}}

\newcommand{\minusmod}[1] 
  {\ensuremath{-}}

\newcommand{\plusmod}[1] 
  {\ensuremath{+}}

\newcommand{\circrangefromto}[3]
  {\rangefromto{#2}{#3}}

\newcommand{\circrangefromopento}[3]
  {\rangefromopento{#2}{#3}}

\newcommand{\circrangeopenfromopento}[3]
  {\rangeopenfromopento{#2}{#3}}

\newcommand{\rangefromto}[2]
 {\ensuremath{[#1:#2]}}

\newcommand{\rangefromopento}[2]
 {\ensuremath{[#1:#2)}}

\newcommand{\rangeopenfromto}[2]
 {\ensuremath{(#1:#2]}}

\newcommand{\rangeopenfromopento}[2]
 {\ensuremath{(#1:#2)}}

\newcommand{\rangeofinterv}[1] 
  {\ensuremath{\rangefromopento{\begofinterv{#1}}{\begofinterv{#1+1}}}}

\newcommand{\rangeofopeninterv}[1] 
  {\ensuremath{\rangefromopento{\begofinterv{#1}}{\Endofinterv{#1}}}}

\newcommand{\absvalof} 
  [2][1] 
  {\delimit{#1}{\lvert}{\rvert}{#2}}

\newcommand{\SetOfRealNumbers}
  {\ensuremath{\mathbb{R}}}

\newcommand{\circcase}
  [1] 
  {\ensuremath{\mbox{}_{\ifthenelse{\equal{#1}{c}}{\mathrm{c}}{}}}}

\newcommand{\TypesetProblemName} 
  [2] 
  [-]
  {\ifthenelse{\equal{#1}{-}}{\MakeTextUppercase{#2}}{\mbox{\MakeTextUppercase{#2}}}}

\newcommand{\ALLINSERTIONSSCAL}
  [2][] 
  {\ifthenelse{\equal{#1}{full}}{\TypesetProblemName{Max\-i\-mal Sums of In\-ser\-tions}}{MSI}\circcase{#2}}

\newcommand{\MSII}
  [1][short] 
  {\ifthenelse{\equal{#1}{full}}{\TypesetProblemName[-]{Max\-i\-mal Linear Sums
  of In\-de\-pend\-ent In\-ser\-tions}}{\TypesetProblemName[box]{MLSII}}}
\newcommand{\MCSII}
  [1][short] 
  {\ifthenelse{\equal{#1}{full}}{\TypesetProblemName[-]{Max\-i\-mal Circular
  Sums of In\-de\-pend\-ent In\-ser\-tions}}{\TypesetProblemName[box]{MCSII}}}
\newcommand{\SSS}[1][short]
  {\ifthenelse{\equal{#1}{full}}{\TypesetProblemName{Sorting a Se\-quence by Scores}}{SSS}}

\newcommand{\circsubseq} 
  [3] 
  {\subseq{#1}{#2}{#3}}

\newcommand{\circsubseqopen}
  [3]
  {\subseqopen{#1}{#2}{#3}}

\newcommand{\circsubopenseq}
  [3]
  {\subopenseq{#1}{#2}{#3}}

\newcommand{\genminsumat}
  [3][] 
  {\unaryfunctionof{\mathcal{B\ifthenelse{\equal{#1}{c}}{C}{L}SS}_{#2}}{#3}}

\newcommand{\genmaxsumat}
  [3][] 
  {\unaryfunctionof{\mathcal{M\ifthenelse{\equal{#1}{c}}{}{}SS}_{#2}}{#3}}

\newcommand{\genmaxsumfrom}
  [3][] 
  {\unaryfunctionof{\mathcal{M\ifthenelse{\equal{#1}{c}}{}{}PS}_{#2}}{#3}}

\newcommand{\genmaxsumof}
  [2][] 
  {\unaryfunctionof{\mathcal{M\ifthenelse{\equal{#1}{c}}{C}{L}S}}{#2}}

\newcommand{\genmaxsumofof}
  [2][] 
  {\unaryfunctionof{\mathcal{M\ifthenelse{\equal{#1}{c}}{C}{L}S}_2}{#2}}

\newcommand{\maxgapat}
  [2] 
  {\unaryfunctionof{\mathcal{MG}_{#1}}{#2}}

\newcommand{\maxcircsumat}
  [2] 
  {\genmaxsumat[c]{#1}{#2}}

\newcommand{\maxcircsumof}
  [1] 
  {\genmaxsumof[c]{#1}}

\newcommand{\maxreachpeak}
  [1] 
  {\ensuremath{Q_{#1}}}

\newcommand{\maxreachpos}
  [1] 
  {\ensuremath{q_{#1}}}

\newcommand{\maxsumat}
  [2] 
  {\genmaxsumat{#1}{#2}}

\newcommand{\minsumat}
  [2] 
  {\genminsumat{#1}{#2}}

\newcommand{\maxsumfrom}
  [2] 
  {\genmaxsumfrom{#1}{#2}}

\newcommand{\maxsumof}
  [1] 
  {\genmaxsumof{#1}}

\newcommand{\maxsumofof}
  [1] 
  {\genmaxsumofof{#1}}

\newcommand{\maxcircsumfrom}
  [2] 
  {\genmaxsumfrom[c]{#1}{#2}}

\newcommand{\peakofinterv}
  [1] 
  {\ensuremath{p_{#1}}}

\newcommand{\seq} 
  [2] 
  [1]
  {\delimit{#1}{\langle}{\rangle}{#2}}

\newcommand{\seqconcat}{\ensuremath{}}

\newcommand{\seqexpl} 
  [2] 
  {\seq{\seqindex{#1}{0}, \ldots, \seqindex{#1}{#2-1}}}

\newcommand{\seqidx}
  [2] 
  {\ensuremath{#1[#2]}}

\newcommand{\seqindex}
  [2] 
  {\seqidx{#1}{#2}}

\newcommand{\seqins} 
  [3][omittedelement] 
  {\ensuremath{#2^{\ifthenelse{\equal{#1}{omittedelement}}{}{(#1 \rightarrow}
  #3\ifthenelse{\equal{#1}{omittedelement}}{}{)}}}}

\newcommand{\shiftseqleft}[2]
  {\binfunctionof{\mathit{shift}}{#1}{#2}}

\newcommand{\sizeofseq} 
  [2] 
  [1]
  {\delimit{#1}{\lvert}{\rvert}{#2}}

\newcommand{\subseq} 
  [3] 
  {\ensuremath{#1[#2:#3]}}

\newcommand{\subseqopen} 
  [3] 
  {\ensuremath{#1[#2:#3)}}

\newcommand{\subopenseq} 
  [3] 
  {\ensuremath{#1(#2:#3]}}

\newcommand{\subopenseqopen} 
  [3] 
  {\ensuremath{#1(#2:#3)}}

\newcommand{\sumof}
  [1] 
  {\unaryfunctionof{\mathit{sum}}{#1}}

\newcommand{\elemSep} 
  {\ensuremath{,\:}}

\renewcommand{\emptyset}{\ensuremath{\varnothing}}

\newcommand{\set}
  [2] 
  [1]
  {\delimit{#1}{\{}{\}}{#2}}

\newcommand{\setconditionseparator}{:}
\newcommand{\setofsuchthat} 
  [3] 
  [1]
  {\set[#1]{#2 \setconditionseparator #3}}

\newtheorem{lemma}{Lemma}

\newtheorem{theorem}{Theorem}

\InsertSubmissionOrSelfInput
{
\newproof{proof}{Proof}
}
{

}

\newcommand{\MyQEDHere}{\InsertSubmissionOrSelfInput{\qed}{\qedhere}}

%% file: macros_over-underbrace_in_tikz.tex
\newcommand{\OverbraceTikZPoints} 
  [4][0pt] 
  {\path (#2) -- (#3) let \p1 = ($(#3) - (#2)$), \n1 = {veclen(\x1,\y1)}
                      in node[pos=0.5,above=#1,sloped] {$\overbrace{\hspace{\n1}}^{#4}$} ; }

\newcommand{\UnderbraceTikZPoints} 
  [4][0pt] 
  {\path (#2) -- (#3) let \p1 = ($(#3) - (#2)$), \n1 = {veclen(\x1,\y1)}
                      in node[pos=0.5,below=#1,sloped] {$\underbrace{\hspace{\n1}}_{#4}$} ; }

%% file: macros_text.tex
\newcommand{\boxedreminder}
  [2][math] 
  {\reminder{\ifthenelse{\equal{#1}{math}}{\boxed}{\fbox}{#2}}}

\newcommand{\definedexpr} 
  [2][emph] 
  {\ifthenelse{\equal{#1}{emph}}
              {\emph{#2}}
              {\ifthenelse{\equal{#1}{strong}}
                          {\textbf{#2}}
                          {\ifthenelse{\equal{#1}{quotes}}
                                      {``#2''}
                                      {#2}%
                          }%
              }%
  }

\def\versionoftext{long}
\newcommand{\InsertShortOrLongText}[2] 
  {\ifthenelse{\equal{\versionoftext}{short}}{#1}{#2}}

\newcommand{\reminder}
  [1] 
  {\textcolor{red}{#1}}

\newcommand{\remindersymbol} 
  {\reminder{\mbox{$\bigstar$\ }}}

\newcommand{\zerowidthreminder}
  [2][l] 
  {\makebox[0pt][#1]{\reminder{\textbf{#2}}}}

%% file: intro.tex
\section{Introduction} \label{SecUnitIntroduction}%

Let a \emph{sequence} of $n$ elements be denoted by $A = \seqexpl{A}{n}$
  and its \emph{size} by $\sizeofseq{A} = n$.
The aim in this paper is to provide efficient algorithms to answer
  certain insertion-related queries on a sequence of numbers.
For a given sequence $A$ of $n$ real numbers,
  a query takes as arguments a real number $x$ and an index $p \in \set{0, \ldots, n}$,
  and returns the ``score'' of sequence \seqins[x]{A}{p},
  the latter being the sequence which results
  from inserting $x$ just after element \seqindex{A}{p-1}.
By the ``score'' of a sequence $B$ of real numbers we mean
  the greatest sum of all elements of any contiguous, possibly empty subsequence $S$ of $B$.
The focus in this paper is on \emph{independent} queries,
  that is, given a fixed sequence $A$,
  we want to answer a number of unrelated queries on the same sequence $A$.

\subsection{Definitions and Results}

For $n \in \mathbb{N}$, we use the additive group modulo $n$ to handle
the indices of elements of sequences of size $n$. In this sense,
$\rangefromto{i}{j}$, where $i \in \{ 0, \ldots, n-1\}$ and $j \in
\set{0,\ldots,n}$, denotes the set of indices $\set{ i, (i+1)\!\mod n, \ldots, j }$. We write
$\rangeopenfromto{i}{j}$ for the set $\rangefromto{i}{j} \setminus \set{ i}$.
Similarly, $\rangefromopento{i}{j}$ stands for the sequence obtained by removing
$j$ from $\rangefromto{i}{j}$. 
If $i=j$, then $\rangeopenfromto{i}{j}$ and $\rangefromopento{i}{j}$ are the empty set. 
Let $\rangefromto{i}{j}$ be such that $\rangefromto{i}{j} \subseteq
\rangefromopento{0}{n}$. In this case, $\subseq{A}{i}{j}$ is a
\emph{subsequence of $A$} constituted by all elements of $A$ indexed by indices
in $\rangefromto{i}{j}$ and taken in the same relative order as they appear in
$A$. We assume that $A[0]$ succeeds $A[n-1]$ in this order so that $\langle
A[i], \ldots, A[n-1], A[0], \ldots, A[j] \rangle$, with $i > j$, is a
subsequence of $A$ defined by $\rangefromto{i}{j}$. Analogously,
$\subopenseq{A}{i}{j}$, $\subseqopen{A}{i}{j}$, and $\subopenseqopen{A}{i}{j}$
are subsequences of $A$. The sequence $\subseq{A}{i}{i-1}$ is a \emph{circular
shift of $A$ up to $i$}. If $i=j$, then $\subopenseq{A}{i}{j}$ and
$\subseqopen{A}{i}{j}$
are the empty sequence $\seq{}$, including if $i=j=n$.

The \emph{concatenation} of sequences $A$ and $B$
  of sizes respectively
  $n$ and $m$ is denoted by
  $A \seqconcat B = \seq{\seqindex{A}{0}, \ldots, \seqindex{A}{n-1}} \seqconcat
  \seq{\seqindex{B}{0}, \ldots, \seqindex{B}{m-1}} = \seq{\seqindex{A}{0},
  \ldots, \seqindex{A}{n-1}, \seqindex{B}{0}, \ldots, \seqindex{B}{m-1}}$. The
  sequence $\subseqopen{A}{0}{p} \seqconcat \seq{x} \seqconcat \subseqopen{A}{p}{n}$ which results from the \emph{insertion}
  of an element $x$ into position $p \in \rangefromto{0}{n}$ of $A$
  is denoted by $\seqins[x]{A}{p}$. We may also use the abbreviation
  $\seqins{A}{p}$ when element $x$ is clear by context.
If $A$ is a sequence of \emph{real numbers},
  then its sum is $\sumof{A} = \sum_{i=0}^{n-1} \seqindex{A}{i}$,
  which equals zero if $A = \seq{}$.
Moreover,  the \emph{maximal linear sum} of $A$,
  denoted by \maxsumof{A}, is the greatest sum of a subsequence $\subseqopen{A}{i}{j}$, with
  $0 \leq i \leq j \leq n$, of $A$. Note that $\maxsumof{A} \geq 0$ due to the
  empty subsequence. \maxsumof{\seqins{A}{p}} can be obtained by simply applying the Kadane's linear
time algorithm (described in {\SecUnitGeneralApproach}) to $\seqins{A}{p}$.
Denoted by \maxcircsumof{A}, the \emph{maximal circular sum} of $A$ is defined
as the greatest sum of a subsequence
$\subseqopen{A}{i}{j}$, with $0 \leq i,j \leq n$, of $A$.
  Note that $\maxcircsumof{A} \geq \maxsumof{A}$ since only subsequences
  $\subseqopen{A}{i}{j}$ with $i \leq j$ are considered in
  \maxsumof{A}.

Let a \emph{query} be a pair $(x, p)$, $x \in \SetOfRealNumbers$, and $p \in
\rangefromto{0}{n}$. We consider in this paper two optimization problems
defined on a sequence $A$ of $n$ real numbers and $m$ queries $(x_0, p_0),
\ldots, (x_{m-1}, p_{m-1})$. The \MSII[full] (\MSII) is defined as the problem
of determining the maximal linear sums of the resulting sequences
$\seqins[x_0]{A}{p_0}, \ldots, \seqins[x_{m-1}]{A}{ p_{m-1} }$. Similarly, the
\MCSII[full] (\MCSII) problem asks for the maximal circular sums of the
resulting sequences. An
extension of Kadane's algorithm to solve the {\MSII} problem takes $\Omega(nm)$
time. In this paper, we give an \Oof{n+m} time algorithm to solve
the {\MSII} problem that handles any number $m$ of queries in a sequence of $n$
numbers, each query concerning an arbitrary insertion position $p$ in the sequence.
A second contribution in this paper is an \Oof{n+m} time algorithm to
solve the {\MCSII} problem. This second algorithm turned out not to be a
straightforward extension of the first one, having demanded several
scattered technical observations. In particular, we show that it is related to
the problem of finding the minimal sum of a subsequence in some cases.

Our algorithms to solve the {\MSII} and {\MCSII} problems are divided in two
phases. In the first phase, a \emph{preprocessing step} is performed by taking
as argument the sequence $A$ and computing the \emph{summary of $A$}, which
consists of an array storing special aggregate information about specific
families of subsequences of $A$. The second phase is a sequence of $m$
\emph{query-answering steps}, each one taking as arguments a query $(x_i, p_i)$,
$i \in \rangefromopento{0}{m}$, and the summary produced by the preprocessing step to compute \maxsumof{\seqins[x] {A} {p}}, in the linear case, or
\maxcircsumof{\seqins[x] {A} {p}}, in the circular case.
The \Oof{n+m} time complexity to solve the {\MSII} and {\MCSII} problems stems from the fact that the preprocessing step takes \Oof{n} time
  and the query-answering one takes \Oof{1} time.

\subsection{Motivation and Applications} \label{SecUnitMotivationAndApplications}%

The algorithms proposed in this paper were motivated by a buffer minimization problem in wireless mesh networks, as follows.
In a radio network,
  interference between nearby transmissions prevents simultaneous communication
  between pairs of nodes which are sufficiently close to each other.
One way to circumvent this problem is to use a \emph{time division multiple access} (TDMA) communication protocol.
In such a protocol,
  the communication in the network proceeds by the successive repetition of a sequence of \emph{transmission rounds},
  in each of which only noninterfering transmissions are allowed to take place.
Such protocols can also be used in the particular case of a \emph{wireless mesh network},
  where each node not only communicates data relevant to itself,
  but also forwards packets sent by other nodes,
  thus enabling communication between distant parts of the network%
  ~\cite{BibAkyildizWangWang2005,BibKlasingMoralesPerennes2008}.
In this case,
  each node stores in a buffer the packets that it must still forward,
  and the optimization of buffer usage in such networks is a current research
  topic%
  ~\cite{BibVieiraRezendeBarbosaFdida2012,BibLeModianoShroff2012}.

In order to analyze the use of buffer space in a given set of $m$ nodes of a
network, we represent a sequence of $n$ transmission rounds by an $m \times n$
matrix $R$, defined as follows:
  $\matrixelem{R}{i}{j} = +1$ if node $i$ receives a packet at round $j$;
  $\matrixelem{R}{i}{j} = -1$ if at round $j$ node $i$ forwards a packet; and
  $\matrixelem{R}{i}{j} = 0$ otherwise. The positive value of
  $\matrixelem{R}{i}{j}$ stands for the additional memory space required by node
  $i$ to store the packet received at round $j$, whereas the negative value accounts for
  the memory space released by node $i$ when a packet is sent. Since the
  same sequence of rounds is successively repeated in the network, it can be
  verified that the greatest number of packets that node $i$ will ever need to
  store simultaneously corresponds to the maximal circular sum of row $i$ of
  $R$. The buffer usage for the whole set of $m$ nodes is then given by
  $\costofmatrix{R} = \sum_{i=0}^{m-1} \maxcircsumof{ R_i }$, where, for
  $i \in \rangefromopento{0}{m}$, $R_i$ denotes the sequence corresponding to the
  $R$'s $i$th row. This leads us to the following NP-Hard problem:
  given a matrix $R$ representing a sequence of transmission rounds,
  find a permutation $R'$ of the columns of $R$ which minimizes
  \costofmatrix{R'}. The algorithms introduced in the present paper provide a valuable tool
  for the development of heuristics to this problem.

The concept of \emph{maximal sum subsequence}
  and its generalizations for two dimensions 
  are currently known to have several other applications in practice,
  for example in Pattern Recognition~\cite{BibBentley1984,
                                           BibAnPeursumLiuVenkatesh2009}, 
  Data Mining~\cite{BibFukudaMorimotoMorishitaTokuyama2001}, 
  Bioinformatics~\cite{BibRuzzoTompa1999,BibLinJiangChao2002,BibCsuros2004}, 
  Health and Environmental Science~\cite{BibFukudaTakaoka2007}, 
  Medicine~\cite{BibThaherTakaoka2010}, 
  and Strategic Planning~\cite{BibLu2010}; 
  consequently, it is plausible that other applications
  of our algorithms be found in the future.

\subsection{Related Work} \label{SecUnitRelatedWork}%

We are not aware of any previous attempt to solve the main problem tackled in
  this paper -- i.e.\ computing \maxsumof{\seqins[x]{A}{p}} and
  \maxcircsumof{\seqins[x]{A}{p}} for any given $x$ and $p$,
  for a fixed sequence $A$. However, there is a large number of works related to
  subsequence sums.
The basic problem of finding a maximal sum subsequence of a sequence $A$ of $n$ numbers
  was given an optimal and very simple solution by Joseph B. Kadane around 1977.
The algorithm, which takes \Oof{n} time,
  was discussed and popularized by Gries~\cite{BibGries1982} and Bentley~\cite{BibBentley1984}.
This one-dimensional problem can be generalized for any number $d$ of dimensions.
The two-dimensional case
  consists in finding a maximal sum submatrix of a given $m \times n$ matrix of numbers,
  and it can be solved in \Oof{m^2 \cdot n} time, with $m \leq n$~\cite{BibBentley1984November}.
Asymptotically slightly faster algorithms do exist
  but are reported not to perform well in practice except for very large inputs%
  ~\cite{BibAnPeursumLiuVenkatesh2009,BibFukudaTakaoka2007}.
More recently, the two-dimensional case has also been explored in the direction of
  convex but not necessarily rectangular
  shapes~\cite{BibThaherTakaoka2010,BibThaherTakaoka2012}. The problem of computing
  \maxsumof{\subseqopen{A}{i}{j}}, for any given $i$ and $j$, $0 \leq i \leq j \leq
  n$, has been tackled in~\cite{BibChenChao2007},
  where a two-phase algorithm consisting of a linear preprocessing time plus
  constant time per query is presented.

Another direction of generalization of the original problem which has been explored is that of
  finding multiple maximal sum subsequences instead of just one.
Finding a set of \emph{all} successive and nonintersecting maximal sum
subsequences of $A$ can be done in \Oof{n}
sequential time~\cite{BibRuzzoTompa1999},
  \Oof{\log n} parallel time in the EREW PRAM model~\cite{BibDaiSu2006}
  and \Oof{\sizeofseq{A}/p} parallel time with $p$ processors in the BSP/CGM model~\cite{BibAlvesCaceresSong2006}.
A different problem,
  concerning the maximization of the \emph{sum} of any set of $k$ nonintersecting subsequences,
  is considered in~\cite{BibCsuros2004}.
A list of the $k$ \emph{possibly intersecting} maximal sum subsequences
  of a given sequence of $n$ numbers can be found in optimal \Oof{n+k} time and
  \Oof{k} space~\cite{BibBrodalJorgensen2007}.
%

With motivations from Bioinformatics, a further direction of research concerns problems
  where one or more measures of subsequences are constrained.
For example, algorithms have been devised that
  compute the greatest sum among all subsequences subject to
  a length lower bound~\cite{BibHuang1994},
  a length upper bound~\cite{BibLinJiangChao2002,BibMu2008},
  both length bounds~\cite{BibLinJiangChao2002}
  or average bounds~\cite{BibChengLiuChao2009}
  (the average of a sequence $A$ being $\sumof{A}/\sizeofseq{A}$).
Optimal algorithms have also been devised for the length constrained versions of
  the multiple maximal sum subsequences~\cite{BibBrodalJorgensen2008}.
%
%

To the best of our knowledge,
  the column permutation problem defined in the previous subsection
  has not yet been considered in the literature. 
The closest related and already studied problem that we know of is
  the following variation of it for only \emph{one row}:
  given a sequence $A$ of $n$ real numbers,
  find a permutation $A'$ of $A$ which minimizes \maxsumof{A'}.
This problem was found to be solvable in \Oof{\log n} time
  in the particular case where $A$ has only two distinct numbers~\cite{BibLiHui1992};
  the same paper also mentions that the case where $A$ may have arbitrary numbers
  can be shown to be strongly NP-hard by reduction from the \TypesetProblemName{3-partition} problem.
Such a reduction has actually been presented recently,
  together with an \Oof{n \log n} algorithm
  which has an approximation factor of 2 for the case of arbitrary input numbers
  and $3/2$ for the case where the input numbers
  are subject to certain restrictions~\cite{BibCorreaFariasSouza2013}.

%
%


\subsection{Structure of the Paper}

The remaining of this paper is structured as follows.
{\SecUnitGeneralApproach} introduces our approach as well as the preprocessing and query-answering algorithms for the linear case.
Our preprocessing and query-answering algorithms for the circular
case are then presented in Section~\ref{SecUnitGenApproachCirc}.
{\SecUnitConclusion} presents our concluding remarks, closing the paper.

%% file: linear.tex
\newcommand{\arrayintervidx}{\arrayname{LRA}}%
\newcommand{\arrayintervidxof}[1]{\arrayindex{\arrayintervidx}{#1}}%
\definenewarray{leftmaxsum}{LMLS} 
\definenewarray{rightmaxsum}{RMLS} 
\newcommand{\newseq}{\seqins{A}{p}}%
%
\newcommand{\result}{\maxsumof{\newseq}}%

\newcommand{\arrayofmaxsumsat}[1]{\arrayindex{\ensuremath{\mathit{MSS}}}{#1}}%
\newcommand{\arrayofmaxsumsfrom}[1]{\arrayindex{\ensuremath{\mathit{MPS}}}{#1}}%

\section{Linear Case} \label{SecUnitGeneralApproach}%

In this section, we consider the \MSII\ problem with respect to a
sequence $A$ of $n \geq 0$ real numbers and $m$ queries $(x_0, p_0), \ldots,
(x_{m-1}, p_{m-1})$. We call the {\em inverse} of $A$, denoted by $A^{-1}$, the
sequence $\langle A[n-1], \ldots, A[0] \rangle$. If $i \in
\rangefromopento{0}{ n }$, $j \in \rangefromto{0}{ n }$, and $h \in
\rangefromto{i}{j}$, then we say that $\subseqopen{A}{i}{h}$ and $\subseqopen{A}{h}{j}$ are respectively a
\emph{prefix} and a \emph{suffix} of $\subseqopen{A}{i}{j}$.
Let the \emph{maximal suffix sum} of $A$ be defined
as $\maxsumat{}{A} = \max
\setofsuchthat{\sumof{\subseqopen{A} {i} {n}}} {i \in
\rangefromto{0}{n}}$ ($\maxsumat{}{A} \geq 0$ since $\sumof{\subseqopen{A}
{n} {n}} = 0$).

\subsection{General Approach}

The general approach to handle the multiple insertions fast is better seen as
the two-phase algorithm mentioned in the Introduction. In this vein, the
\emph{summary} of a sequence, as depicted in {\tableNoncirc}, consists of some
scalars and an array storing information about two groups of subsequences of $A$.
Informally speaking, one group contains subsequences of $\subseqopen{A}{0}{i}$
and $\subseqopen{A}{i}{n}$, for every $i \in \rangefromto{0}{n}$. These
are the subsequences which are candidates to be a maximum sum subsequence of
$\seqins[x]{A}{i}$ not including $x$. The other group is
composed by the subsequences of $A$ defined by a concatenation of a suffix of $\subseqopen{ A }{
0 }{i}$ and a prefix of $\subseqopen{ A }{ i }{n}$. Note that the sum of a
prefix of $\subseqopen{ A }{ i }{n}$ is also the sum of a suffix of $\subseqopen{A^{-1}}{0}{n-i}$.

The computation of the summary of $A$ given in {\tableNoncirc} and illustrated
in Fig.~\ref{FigGenApproachNoncircBeforeIns} is performed as follows. For the
purpose of capturing subsequences of $A$ in the general expression used in the
query-answering step (which is useful in the analysis of the circular case in
Section~\ref{SecUnitGenApproachCirc}), let $q \in \rangefromopento{ 0 }{ n }$,
$r \in \rangefromto{ 0 }{ n }$, $q \leq r$, and $(x, p)$ be a query, $p \in
\rangefromto{q}{r}$. The maximal linear sum of
$\subseq{\seqins[x]{A}{p}}{q}{r}$ is given by
\begin{equation}
\maxsumof{\subseq{\seqins[x]{A}{p}}{q}{r}} = \max \{ \maxsumof{ \subseqopen{
A }{ q }{ p } }\elemSep \maxsumof{ \subseqopen{ A }{ p }{ r }
}\elemSep \maxsumat{}{\subseqopen{A}{q}{p}} +x+
\maxsumat{}{\subseqopen{A^{-1}}{n-r}{n-p}} \}.
\label{eq:noncircposx}
\end{equation}
is computed. For instance, the maximum linear sum of
the sequence $\seqins{A}{p}$ shown in Fig.~\ref{FigGenApproachNoncircAfterIns} is given by
the subsequence including $x = 12$ depicted in the figure, and its value is
derived from~\eqref{eq:noncircposx} as
\[
\maxsumat{}{\subseqopen{A}{0}{8}}+12+\maxsumat{}{\subseqopen{A^{-1}}{0}{8}} =
37.
\]
Expression~\eqref{eq:noncircposx} is a special case of an
algorithm presented in~\cite{Jeuring1991}.

\begin{table}[htb]
\begin{center}
$\begin{array}{rl} \toprule
\multicolumn{2}{c}{\text{Scalars defining the maximum sum subsequence}} \\ \midrule
i^* &= \text{ first index} \\
j^* &= \text{ last index} \\
MAX &= \maxsumof{\subseqopen{A}{q}{r}} = \sumof{\subseqopen{A}{i^*}{j^*}} \\ \midrule\midrule
\multicolumn{2}{c}{\text{Array with indices } i \in \rangefromto{ q }{ r }} \\ \midrule
S[i] &= \begin{cases} \max \set{ \maxsumof{\subseqopen{A}{q}{i}},
\maxsumof{\subseqopen{A}{i}{r}} }, &
\text{if } i \in \rangefromopento{i^*}{j^*} \\
\maxsumat{}{\subseqopen{A}{q}{i}} + \maxsumat{}{\subseqopen{A^{-1}}{n-r}{n-i}},
& \text{otherwise} \end{cases} \\ \bottomrule
\end{array}$
\end{center}
\caption{Summary of $\subseqopen{A}{q}{r}$, assuming that $q \in
\rangefromopento{ 0 }{ n }$, $r \in
\rangefromto{ 0 }{ n }$, and $q \leq r$.}
\label{tableNoncirc}
\end{table}

In the sequel, we discuss how to perform the preprocessing step in \Oof{n}
time, and how the query-answering step can be done in constant time by
considering the information in the summary of $A$.

\edef\A{{2, -7, 4, -25, 12, -1, -8, 14, 1, -6, -3, 5, 11, -18, 8, 10}}
\edef\MLSQ {{0,  2, 2, 4, 4, 12, 12, 12, 17, 18, 18, 18, 18,  25, 25,
25, 25}}
\edef\MLSR {{25,  25, 25, 25, 25, 22, 22, 22, 18, 18, 18, 18, 18, 18, 18,
10, 0}}
\edef\R {{25,  25, 25, 25, 25, 22, 22, 22, 18, 18, 18, 18, 18, 25, 25,
25, 25}}
\edef\SL {{0, 2, 0, 4, 0, 12, 12, 12, 17, 18, 18, 18, 18, 25, 7, 15, 25}}
\edef\SR {{2, 0, 4, 0, 25, 13, 14, 22,  8,  7, 13, 16, 11,  0, 18, 10,  0}}
\edef\S {{ 2, 2, 4, 4, 25, 25, 25, 25, 25, 25, 25, 25, 25, 25, 25, 25, 25}}
\edef\SUMMARY {{ 2, 2, 4, 4, 25, 22, 22, 22, 17, 18, 18, 18, 18, 25, 25, 25,
25}}
\edef\arrayNumElem{16}
\EdefAsResultOf{\lastIdx}{int(\arrayNumElem-1)}
\edef\IntervalColors{{"gray!20","cyan","violet!50","orange","green","pink","brown!80","yellow","red","teal!80"}}%
\edef\K {{0, 0, 0, 0, 1, 1, 1, 1, 1, 1, 1, 1, 1, 0, 0, 0}}
\edef\KI {{0,  0, 0,   0,  0, 0,  0, 0, 0,  0,  0, 1,  1,  2, 2,  3}}

\begin{figure}[htb]
\edef\spaceBtwIndexArray{\elemYSide/6}%
\edef\ExtraSpaceBtwArrayAndItsLabel{\elemXSide/9}%
%
{\centering
        \begin{subfigure}[A sequence and its summary ($MAX = 25$).]{
        \label{FigGenApproachNoncircBeforeIns}
\begin{tikzpicture}
\node at (-\elemXSide, \arrayYPos + \elemYSide/2) {$i$};
\foreach \x in {0,...,\arrayNumElem} {
								  \node at (\x*\elemXSide+\elemXSide/2, \arrayYPos + \elemYSide/2)
								  {\ifnum \x = 4
								  	$i^*$
								   \else \ifnum \x = 13
								    $j^*$
								   \else
								   	\x
								   \fi
								   \fi}; }
\EdefAsResultOf{\arrayYPos}{\arrayYPos-\elemYSide}

\node at (-\elemXSide, \arrayYPos + \elemYSide/2) {$\seqindex{A}{i}$};
\foreach \x in {0,...,\lastIdx} { \EdefAsResultOf{\number}{int(\A[\x])}
                                  \EdefAsResultOf{\fillOfSubseq}{\IntervalColors[\K[\x]]}
                                  \node[rectangle,draw,minimum width=\elemXSide cm,minimum height=\elemYSide cm,fill=\fillOfSubseq] at
								  (\x*\elemXSide+\elemXSide/2, \arrayYPos + \elemYSide/2) {\number};
								  }
\node[rectangle,draw,minimum width=\elemXSide cm,minimum height=\elemYSide
cm] at (\arrayNumElem*\elemXSide+\elemXSide/2, \arrayYPos + \elemYSide/2) {--};
\EdefAsResultOf{\arrayYPos}{\arrayYPos-\elemYSide}

\node at (-\elemXSide, \arrayYPos + \elemYSide/2)
{$S[i]$};
\foreach \x in {0,...,\arrayNumElem} {\EdefAsResultOf{\number}{int(\SUMMARY[\x])}
								  \node[rectangle,minimum width=\elemXSide cm,minimum height=\elemYSide
								  cm] at (\x*\elemXSide+\elemXSide/2, \arrayYPos + \elemYSide/2) {\number}; }
\draw (0, \elemYSide) -- (0, \arrayYPos);
\draw (-2*\elemXSide, 0) -- (0, 0);
\draw (-2*\elemXSide, -\elemYSide) -- (0, -\elemYSide);
\end{tikzpicture}}
        \end{subfigure}\quad\quad
\EdefAsResultOf{\arrayYPos}{\arrayYPos - 2*\elemYSide - 0.4}
        \begin{subfigure}[Sequence resulting from the insertion of $x =
        12$ at position $p = 8$. Subsequences $\seq{12, -1, -8, 14}$ and
        $\seq{11,5,-3,-6,1}$ are maximum sum suffixes of, respectively,
        $\subseqopen{A} {0} {p}$ and $\subseqopen{A^{-1}} {0} {n-p}$.]{
        \label{FigGenApproachNoncircAfterIns}
\begin{tikzpicture}
\input{FigGenApproachNoncircAfterIns.tex}
\end{tikzpicture}}
        \end{subfigure}
\par%
}
\caption{Insertion in the linear case.
        }%
\label{FigGenApproachNoncirc}%
\end{figure}

\subsection{Preprocessing Step}

The computation of the entries of the summary of $A$ is accomplished in
Alg.~\ref{alg:lsumms} through sweeps of $A$ having in mind the fact that the prefix of a maximum sum
subsequence must be nonnegative. The first part, encapsulated in
Alg.~\ref{alg:lsumm}, is a version of Kadane's algorithm devoted to
set a summary (as a global variable) in such a way that, after the call
\CALL{MSS}{A, q, r}, $\subseqopen{A}{i^*}{j^*}$ is a maximum sum subsequence
with $MAX = \sumof{\subseqopen{A}{i^*}{j^*}}$, $S[q] = 0$, and
\[
S[j] = \max \set{0 \elemSep \seqindex{A}{j-1} + S[j-1]} =
\maxsumat{}{\subseqopen{A}{q}{j}}, \text{ for all } j \in
\rangeopenfromto{q}{r}.
\]
The application of Alg.~\ref{alg:lsumm} to the sequence in
Fig.~\ref{FigGenApproachNoncirc} with $q=0$ and $r=16$ produces
\begin{center}
\begin{tabular}{c|ccccccccccccccccc}
$j$ & 0 & 1 & 2 & 3 & 4 & 5 & 6 & 7 & 8 & 9 & 10 & 11 & 12 & 13 & 14 & 15 & 16
\\
\hline $S[j]$ & 0 & 2 & 0 & 4 & 0 & 12 & 11 & 3 & 17 & 18 & 12 & 9 & 14 & 25
& 7 & 15 & 25
\end{tabular}
\end{center}
as the state of $S$, $i^* = 4$, $j^* = 13$, and $MAX = 25$.

\begin{algorithm}[htbp]
\caption{\protect\CALL{MSS}{A, q, r}: Computation
of maximum sum subsequence of $\subseqopen{A}{q}{r}$, $q < r$.}
\begin{algorithmic}[1]
\State $i,i^*, j^* \gets q$ \label{lin:partIbeg}
\State $MAX,S[q] \gets 0$
\For{$j \in \rangeopenfromto{q}{r}$}
	\State $x \gets \seqindex{A}{j-1}+S[j-1]$
	\If{$x < 0$}
		\State $S[j] \gets 0$
		\State $i \gets j$
	\Else
		\State $S[j] \gets x$
		\If {$x > MAX$}
			\State $MAX \gets x$
			\State $j^* \gets j$
			\State $i^* \gets i$
		\EndIf
	\EndIf
\EndFor \label{lin:partIend}
\end{algorithmic}
\label{alg:lsumm}
\end{algorithm}

The second part of Alg.~\ref{alg:lsumms} starts at line~\ref{lin:partIIbeg}
and finishes at line~\ref{lin:partIIend}. In this part, maximum suffix sums of
$A^{-1}$ are computed by traversing $\subseqopen{A}{q}{r}$ in descending order,
leaving $S$ in the state
\[
S[j] = \maxsumat{}{\subseqopen{A}{q}{j}} +
\maxsumat{}{\subseqopen{A^{-1}}{n-r}{n-j}}.
\]
This state for the sequence in Fig~\ref{FigGenApproachNoncirc} is given by
\begin{center}
\begin{tabular}{c|ccccccccccccccccc}
$j$ & 0 & 1 & 2 & 3 & 4 & 5 & 6 & 7 & 8 & 9 & 10 & 11 & 12 & 13 & 14 & 15 & 16
\\
\hline $S[j]$ & 2 & 2 & 4 & 4 & 25 & 25 & 25 & 25 & 25 & 25 & 25 & 25 & 25 & 25
& 25 & 25 & 25
\end{tabular}
\end{center}
Only the maximum sum subsequence $\subseqopen{A}{i^*}{j^*}$ is considered in the
next two parts. In lines~\ref{lin:partIIIbeg}--\ref{lin:partIIIend}, the state
of $S$ is modified in order to satisfy $S[j] = \maxsumof{\subseqopen{A}{q}{j}}$,
for all $j \in \rangefromopento{i^*}{j^*}$.
The call \CALL{MLS}{A, i, j} returns $\maxsumof{\subseqopen{A}{i}{j}}$, computed in linear time through Kadane's algorithm (the detailed
description of \CALL{MLS}{A, i, j} is omitted since it is obtained from \CALL{MSS}{A, i, j} by
simply replacing the global summary by local scalars and returning the local
variable that replaces $MAX$). Considering again the sequence in
Fig.~\ref{FigGenApproachNoncirc}, array $S$ is modified so that the elements for
$j \in \rangefromopento{i^*}{j^*}$ become
\begin{center}
\begin{tabular}{c|ccccccccc}
$j$ & 4 & 5 & 6 & 7 & 8 & 9 & 10 & 11 & 12
\\
\hline $S[j]$ & 4 & 12 & 12 & 12 & 17 & 18 & 18 & 18 & 18
\end{tabular}
\end{center}
Finally, $\maxsumof{\subseqopen{A}{i}{r}}$ is computed in the last part,
lines~\ref{lin:partIVbeg}--\ref{lin:partIVend}.
We can therefore state the
following result.

\begin{theorem}
Algorithm \CALL{summarize}{A,q,r} computes the summary of
$\subseqopen{A}{q}{r}$ in Tab.~\ref{tableNoncirc}, $q \in
\rangefromopento{ 0 }{ n }$, $r \in
\rangefromto{ 0 }{ n }$, and $q < r$, in linear
time.
\end{theorem}

\begin{algorithm}[htbp]
\caption{\protect\CALL{summarize}{A, q, r}: Computation
of the summary of $\subseqopen{A}{q}{r}$, $q < r$.}
\begin{algorithmic}[1]
\State \CALL{MSS}{A, q, r}
\State $b \gets 0$ \label{lin:partIIbeg}
\For{$j \in \rangeopenfromto{q}{r}$}
	\State $b \gets \max\set{0, \seqindex{A}{r-j+q}+b}$
	\State $S[r-j+q] \gets S[r-j+q]+b$
\EndFor \label{lin:partIIend}
\Statex
\State $S[i^*] \gets \CALL{MLS}{A, q, i^*}$ \label{lin:partIIIbeg}
\State $f \gets A[i^*]$
\For{$j \in \rangefromopento{i^*+1}{j^*}$}
	\State $S[j] \gets \max\set{ f, S[j-1]}$
	\State $f \gets f + A[j]$
\EndFor \label{lin:partIIIend}
\Statex
\State $B \gets \CALL{MLS}{A^{-1}, n-r, n-j^*}$ \label{lin:partIVbeg}
\State $b \gets A[j^*-1]$
\State $S[j^*-1] \gets \max\set{S[j^*-1],b,B}$
\For{$i \in \rangeopenfromto{i^*+1}{j^*}$}
	\State $b \gets b + A[j^*-i+i^*]$
	\State $B \gets \max\set{ b, B}$
	\State $S[j^*-i+i^*] \gets \max\set{ S[j^*-i+i^*], B}$
\EndFor \label{lin:partIVend}
\end{algorithmic}
\label{alg:lsumms}
\end{algorithm}

\subsection{Query-answering Step}

The maximal linear sum of the sequence
$\seqins[x]{A}{p}$, $x \in \SetOfRealNumbers$, and $p \in
  \rangefromto{0}{n}$, is determined based on a property of the maximum
  subsequence sum that is captured by the summary in Tab.~\ref{tableNoncirc} as
  follows.

\begin{lemma}
For all $q \in \rangefromopento{ 0 }{ n }$, $r \in
\rangefromto{ 0 }{ n }$, and $i \in \rangefromopento{q}{r}$,
\[
MAX = \left\{ \begin{array}{ll}
\maxsumat{}{\subseqopen{A}{q}{i}} + \maxsumat{}{\subseqopen{A^{-1}}{n-r}{n-i}}, &
\text{if } i \in \rangefromopento{i^*}{j^*} \\
\max \{ \maxsumof{\subseqopen{A}{q}{i}},
\maxsumof{\subseqopen{A}{i}{r}} \}, &
\text{otherwise.} \end{array} \right.
\]
\label{lem:oneismax}
\end{lemma}

\begin{proof}
If
$i \in \rangefromopento{i^*}{j^*}$, let $\hat \i = \argmax_{j \in
\rangefromto{q}{i}} \set{\sumof{\subseqopen{A} {j} {i}}}$ and $\hat \j =
\argmax_{j \in \rangefromto{i}{r}} \set{\sumof{\subseqopen{A} {i} {j}}}$. Notice
that $\maxsumat{}{\subseqopen{A}{q}{i}} = \sumof{\subseqopen{A} {\hat \i} {i}}$,
that a suffix of $\subseqopen{A^{-1}}
{n-r} {n-i}$ is a prefix of $\subseqopen{A} {i} {r}$, and that
$\maxsumat{}{\subseqopen{A^{-1}}{n-r}{n-i}} = \sumof{\subseqopen{A} {i} {\hat \j}}$.
We claim that $\sumof{\subseqopen{A}{\hat \i}{i}} =
\sumof{\subseqopen{A}{i^*}{i}}$ and $\sumof{\subseqopen{A}{i}{\hat \j}} =
\sumof{\subseqopen{A}{i}{j^*}}$. For the former claim, we first observe that
either $\subseqopen{A}{i^*}{\hat \i}$ is a prefix of $\subseqopen{A}{i^*}{j^*}$
or $\subseqopen{A}{\hat \i}{i^*}$ is a prefix of
$\subseqopen{A}{\hat \i}{\hat \j}$. We conclude that
$\sumof{\subseqopen{A}{i^*}{\hat \i}} = \sumof{\subseqopen{A}{\hat \i}{i^*}} = 0$ because, otherwise, the
maximality of either $\sumof{\subseqopen{A}{i^*}{j^*}}$ or
$\sumof{\subseqopen{A}{\hat \i}{\hat \j}}$ would be violated. The reasoning
for the latter claim is similar, by considering suffixes of
$\subseqopen{A}{i^*}{j^*}$ or $\subseqopen{A}{\hat \i}{\hat \j}$. Therefore, $MAX =
\sumof{\subseqopen{A}{i^*}{i}} + \sumof{\subseqopen{A}{i}{j^*}} = \sumof{\subseqopen{A}{\hat
\i}{i}} + \sumof{\subseqopen{A}{i}{\hat \j}}$. The case $i \not\in \rangefromopento{i^*}{j^*}$ stems directly from
$\subseqopen{A}{i^*}{j^*} \subseteq \subseqopen{A}{q}{i}$ or
$\subseqopen{A}{i^*}{j^*} \subseteq \subseqopen{A}{i}{r}$.
\MyQEDHere
\end{proof}

We finally get the main result for the linear case directly
from~\eqref{eq:noncircposx} and the definition of the summary in
Tab.~\ref{tableNoncirc}.

\begin{theorem}
Given a summary as defined in {\tableNoncirc} with $q=0$ and $r=n$,
  the query-answering algorithm computes \maxsumof{ \seqins[x]{A}{p} } in \Oof{1} worst-case time
  for any sequence $A$, $x \in \SetOfRealNumbers$, and $p \in
  \rangefromto{0}{n}$ as
\[
\maxsumof{\seqins{A}{p}} = \left\{ \begin{array}{ll}
\max\set{S[p], x + MAX}, &
\text{if } p \in \rangefromopento{i^*}{j^*} \\
\max\set{ MAX, x + S[p] }, &
\text{otherwise.} \end{array} \right.
\]
\end{theorem}

\begin{proof}
By Lemma~\ref{lem:oneismax}, the equivalence between the expression
for $\maxsumof{\seqins{A}{p}}$ in the theorem and~\eqref{eq:noncircposx} with
$q=0$ and $r=n$ is straightforward.
\MyQEDHere
\end{proof}

%% file: FigGenApproachNoncircAfterIns.tex
\edef\arrayNumElem{17}
\EdefAsResultOf{\lastIdx}{int(\arrayNumElem-1)}
\drawEmptyArray
\putLabelOfArray{$\newseq$}

\edef\Ap{{2, -7, 4, -25, 12, -1, -8, 14, 12, 1, -6, -3, 5, 11, -18, 8, 10}}
\edef\Kp {{0, 0, 0, 0, 1, 1, 1, 1, 2, 1, 1, 1, 1, 1, 0, 0, 0}}

\foreach \x in {0,...,\lastIdx} { \EdefAsResultOf{\number}{int(\Ap[\x])}
                               	\EdefAsResultOf{\fillOfSubseq}{\IntervalColors[\Kp[\x]]}
                                  \drawSubseq{\x}{\x}{\number}{}
                                }

\foreach \x in {0,...,\lastIdx} { \EdefAsResultOf{\ms}{int(\x)}
                                  \putNodeInfo{\x}{}{\ms}
                                }

\underbraceNodes{0} {7}  {\subseqopen{A} {0} {p}}
\underbraceNodes{8} {8}  {x}
\underbraceNodes{9} {16} {\subseqopen{A} {p} {n}}

%% file: circular.tex
\section{Circular Case} \label{SecUnitGenApproachCirc}%
\definenewarray{imaxpfsum}{\arrayrightmaxsum}

\newcommand{\DefOfArrayoriginmax}[1]
  {\ensuremath{#1 \minusmod{n}
   \max \setofsuchthat{ j \in \rangefromopento{ 0 }{ n } }
                      { \sumof{ \circsubseq{ A }{ #1 \minusmod{n} j }{ #1 } }
                        =
                        \maxcircsumat{\subseq{A}{#1+1}{#1}}{#1} }}}%
\newcommand{\arraymaxcsum}{\ensuremath{\mathit{MSS}}}%
\newcommand{\arraymaxcsumof}[1]{\arrayindex{\arraymaxcsum}{#1}}%
\newcommand{\arraymaxcpsum}{\ensuremath{\mathit{MPS}}}%
\newcommand{\arraymaxcpsumof}[1]{\arrayindex{\arraymaxcpsum}{#1}}%

\newcommand{\InterQueueElemSpace}{\ }%

We describe in this section our two-phase algorithm for the
{\MCSII} problem. As an additional notation, we write $\bar A$ to
denote the \emph{complement} of $A$, which is the sequence
$\seq{-\seqindex{A}{n-1}, \ldots, -\seqindex{A}{0}}$.
  Note that $\bar A[n-i-1]=-A[i]$, for all $i \in \rangefromopento{0}{n}$, and
  $\bar{\bar A} = A$, as illustrated in Fig.~\ref{FigGenApproachCirc}.
  Also, it is worth noting that $\subseqopen{A}{i^*}{j^*}$ is a maximum
  (minimum) sum subsequence of $A$ if and only if $\subseqopen{\bar
  A}{n-j^*}{n-i^*}$ is a minimum (maximum) sum subsequence of $\bar A$.

\subsection{General Approach}

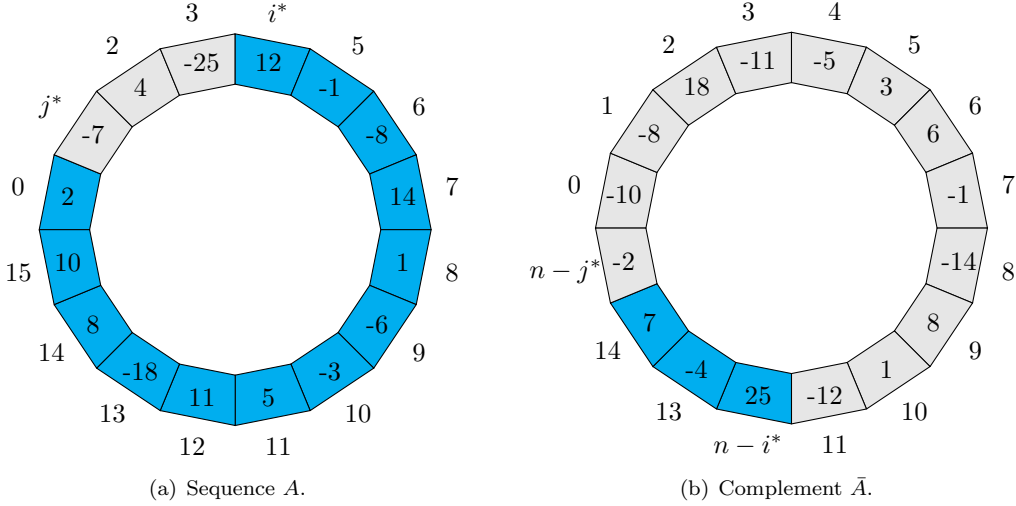
\begin{figure}[htb]

{\centering
        \begin{subfigure}[Sequence $A$.]{
        \label{FigGenApproachCircBeforeIns}
\begin{tikzpicture}
\input{FigGenApproachCircBeforeIns.tex}
\end{tikzpicture}}
        \end{subfigure}\quad\quad
        \begin{subfigure}[Complement $\bar A$.]{
        \label{FigGenApproachCircCompBeforeIns}
\begin{tikzpicture}
\input{FigGenApproachCircCompBeforeIns.tex}
\end{tikzpicture}}
        \end{subfigure}
\par%
}

\caption{A sequence and its complement. In both cases, maximum
sum subsequences are highlighted.}%
\label{FigGenApproachCirc}%
\end{figure}

\begin{figure}[htb]
{\centering
\begin{tikzpicture}
\input{FigGenApproachCircAfterIns.tex}
\end{tikzpicture}
\par%
}
\caption{Sequence resulting from the insertion of $x =
        28$ at position $p = 3$ of the sequence in
        Fig.~\ref{FigGenApproachCircBeforeIns}. Subsequence $\seq{-25}$
        corresponds to the maximum sum subsequence $\seq{25}$ of the complement
        in Fig.~\ref{FigGenApproachCircCompBeforeIns}.}%
\label{FigGenApproachCircAfterIns}
\end{figure}
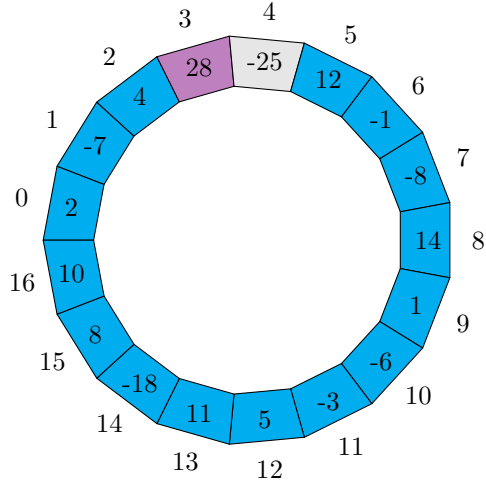

We start the discussion of the circular case by stating the circular version
of~\eqref{eq:noncircposx}. For this purpose, we need to take into account
the sequences of the type $\circsubseq{ A }{ j }{ i }$, with $i \in
\rangefromopento{0}{p}$ and $j \in \rangefromopento{p}{n}$, along with the
subsequences of $A$ considered in~\eqref{eq:noncircposx}.
Hence, we write the maximum circular sum of $\newseq$ as
\begin{equation}
\maxcircsumof{\newseq} = \max \set{ \maxsumof{ \circsubseq{ A
}{ p }{ p-1 } }\elemSep \sumof{A}+x+\maxsumof{\circsubseq{ \bar A }{ n-p }{
n-p-1 }} }.
\label{eq:circposx}
\end{equation}
Inspired by the linear
  case, the value of $\maxcircsumof{\newseq}$ is obtained as the maximum of
  two terms. The first term of~\eqref{eq:circposx} corresponds to the
  maximum sum of the subsequences of the circular shift of $A$ up to $p$,
  which clearly encompass exactly the subsequences of $\newseq$ not including
  $x$. Taking the situation in Fig.~\ref{FigGenApproachCirc} as an example, we observe
  that the maximum sum subsequence of $\circsubseq{ A
}{ 3 }{ 2 }$ is $\seq{ 12, -1, \ldots, 10, 2 }$. Slightly trickier is the fact
that the subsequences including $x$ are considered in the second term. The idea
is to determine the maximum sum of such subsequences as the difference of
$\sumof{A}$ and the minimum subsequence sum of the circular shift of $A$ up to
$p$. This subsequence sum is given by the maximum subsequence sum of
the appropriate circular shift of $\bar A$.
For the example in Fig.~\ref{FigGenApproachCircAfterIns}, we have
\[
\sumof{A} + 28 + \maxsumof{\subseq{\bar A}{13}{12}} = 52.
\]
The central information employed in~\eqref{eq:circposx} is the
maximum subsequence sum of circular shifts of a sequence. We discuss next
that this information is summarized in Tab.~\ref{tableCirc}. For the sake of
illustration, the arrays in the summaries of the sequences in
Fig.~\ref{FigGenApproachCirc} are given by

\vspace{0.2cm}

\edef\arrayNumElem{16}
\EdefAsResultOf{\lastIdx}{int(\arrayNumElem-1)}
\EdefAsResultOf{\antilastIdx}{\lastIdx-1}
\edef\SUMMARY {{25, 24,  24,  24, 27, 24, 24, 24, 20, 20, 20, 20, 20, 25, 25,
25}}
\edef\bSUMMARY{{26, 26, 26, 21,  21, 21, 21, 21, 25, 25, 25, 28, 28, 25, 25,
26}}
\edef\KS {{1, 0, 0, 0, 1, 1, 1, 1, 1, 1, 1, 1, 1, 1, 1, 1}}
\edef\bKS {{0, 1, 1, 1, 0, 0, 0, 0, 0, 0, 0, 0, 0, 0, 0, 0}}

{\centering
\begin{tikzpicture}
\node at (-\elemXSide, \arrayYPos + \elemYSide/2) {$i$};
\foreach \x in {0,...,\lastIdx} {
								  \node at (\x*\elemXSide+\elemXSide/2, \arrayYPos + \elemYSide/2)
								  {\ifnum \x = 4
								  	$i^*$
								   \else \ifnum \x = 1
								    $j^*$
								   \else \ifnum \x = 12
								    $n-i^*$
								   \else \ifnum \x = 15
								    $n-j^*$
								   \else
								   	\x
								   \fi
								   \fi
								   \fi
								   \fi}; }
\EdefAsResultOf{\arrayYPos}{\arrayYPos-\elemYSide}

\node at (-\elemXSide, \arrayYPos + \elemYSide/2)
{$S_A[i]$};
\foreach \x in {0,...,\lastIdx} {\EdefAsResultOf{\number}{int(\SUMMARY[\x])}
                                  \EdefAsResultOf{\fillOfSubseq}{\IntervalColors[\KS[\x]]}
                                  \node[rectangle,draw=\fillOfSubseq,minimum
                                  width=\elemXSide cm,minimum height=\elemYSide cm,fill=\fillOfSubseq] at (\x*\elemXSide+\elemXSide/2, \arrayYPos +
								  \elemYSide/2) {\number}; }
\EdefAsResultOf{\arrayYPos}{\arrayYPos-\elemYSide}

\node at (-\elemXSide, \arrayYPos + \elemYSide/2)
{$S_{\bar A}[i]$};
\foreach \x in {0,...,\lastIdx} {\EdefAsResultOf{\number}{int(\bSUMMARY[\x])}
                                  \EdefAsResultOf{\fillOfSubseq}{\IntervalColors[\bKS[\arrayNumElem-\x-1]]}
                                  \node[rectangle,draw=\fillOfSubseq,minimum
                                  width=\elemXSide cm,minimum height=\elemYSide cm,fill=\fillOfSubseq] at (\x*\elemXSide+\elemXSide/2, \arrayYPos +
								  \elemYSide/2) {\number}; }
\draw (0, \elemYSide) -- (0, \arrayYPos - \elemYSide/2);
\draw (-2*\elemXSide, 0) -- (\arrayNumElem*\elemXSide, 0);
\draw (-2*\elemXSide, -\elemYSide) -- (\arrayNumElem*\elemXSide, -\elemYSide);
\end{tikzpicture}}

\vspace{0.2cm}

\edef\Kc {{1, 0, 0, 0, 1, 1, 1, 1, 1, 1, 1, 1, 1, 1, 1, 1}}

\begin{table}[htb]
\begin{center}
$\begin{array}{rl} \toprule
\multicolumn{2}{c}{\text{Scalars defining the maximum sum subsequence}} \\ \midrule
i^* &= \text{ first index} \\
j^* &= \text{ last index} \\
MAX &= \maxcircsumof{A} = \sumof{\subseqopen{A}{i^*}{j^*}} \\ \midrule \midrule
\multicolumn{2}{c}{\text{Array with indices } i \in \rangefromopento{ 0 }{ n
}}\\
\midrule S_A[i] &= \begin{cases}
\max\left\{
  \maxsumof{\subseqopen{A}{i^*}{i}}, \maxsumof{\subseqopen{A}{i}{j^*}},
  \maxsumof{\subseqopen{A}{j^*}{i^*}},\right. \\
\ \ \ \ \ \ \ \left.
\maxsumof{\subseqopen{A}{i^*}{i}}+\maxsumof{\subseqopen{A}{i}{j^*}}+
\sumof{\subseqopen{A}{j^*}{i^*}}\right\}, & \multirow{-2}{*}{if $i \in
\rangefromopento{i^*}{j^*}$} \\
\sumof{A}+S_{\bar A}[n-i-1], & \text{otherwise}
\end{cases} \\ \bottomrule
\end{array}$
\end{center}
\caption{Summary of $A$ in the circular case.}
\label{tableCirc}
\end{table}

\subsection{Preprocessing Step}%
  \label{SecUnitCircPreproc}%

The computation of the array $S$ for $A$ and $\bar A$ as defined in
Tab.~\ref{tableCirc} can be accomplished by using appropriate slight
modifications of the procedures employed in the linear case. The first thing to do is to compute $MAX$, $i^*$, and $j^*$. In what
        follows, we do it by means of Alg.~\ref{alg:maxsums}, which we derived
        independently, is simple and makes this paper self-contained.
        However, it should be noted that our algorithm is in essence
        very similar to one by Mu~\cite{BibMu2008}, despite much notational
        difference, in part due to the latter having been couched in
        the functional programming paradigm. Moreover, since the circular subsequences of $A$ coincide with the
  linear subsequences of $AA$ with size at most $|A|$, and since the
  length constrained maximum sum problem can be solved in linear time
  by either Lin et al.'s algorithm\cite{BibLinJiangChao2002} or Mu's
        algorithm~\cite{BibMu2008} (see
  Sect.~\ref{SecUnitRelatedWork}), then these algorithms could also be used
  instead of Alg.~\ref{alg:maxsums} to compute $MAX$, $i^*$, and $j^*$.

Alg.~\ref{alg:maxsums} begins with an initially empty list $L$ and then performs
$n$ successive insertions in this list. Each $i$th insertion, for all $i \in
\rangefromopento{0}{n}$ in ascending order, is performed with calls
\CALL{accumulate}{L, A[i]} and \CALL{append}{L, A[i]}, which have the
following specifications: \CALL{accumulate}{L, A[i]}
first adds \seqidx{A}{i} to all current $L$'s members and then removes, from
$L$, all members that are strictly smaller than $A[i]$; \CALL{append}{L, A[i]}
appends \seqidx{A}{i} to $L$ with rank $i$. It turns out that the members that
remain in $L$ after any of these insertions are sorted in ascending order of
rank and non-ascending order of value. Moreover, the members of $L$ after the
$i$th insertion are suffix sums $s_j = \sum_{k \in
\rangefromto{j}{i}} \seqidx{A}{k}$, $j \in \rangefromto{0}{i}$, and
the following properties hold: (P1) the maximum of these suffix sums is
in $L$ and (P2) if $s_{j_1}$, $j_1 < i$, is a member of $L$, then $\max\{s_j
\mid j \in \rangeopenfromto{j_1}{i}\}$ is also a member of $L$. As an illustration of such
a behavior, the members of $L$ just after the $n$ insertions performed at
line~\ref{lin:firstpush} for the sequence in Fig.~\ref{FigGenApproachCircBeforeIns} is
\[
\left[
{
\sum_{k \in \rangefromto{4}{15}} \seqidx{A}{k} = 25,
\sum_{k \in \rangefromto{7}{15}} \seqidx{A}{k} = 22,
\sum_{k \in \rangefromto{14}{15}} \seqidx{A}{k} = 18,
\seqidx{A}{15} = 10
}
\right].
\]
The rank of the first member is 4, of the second, 7, and so on. Notice also
that the member $\sum_{k \in \rangefromto{2}{15}} \seqidx{A}{k} = -9$, of rank
2, is not in this state because it has been removed during the insertion of $\sum_{k \in \rangefromto{4}{15}}
\seqidx{A}{i} = 25$, of rank 4.

Clearly,
the greatest value (and smallest rank) $L$'s member in the state left by the $n$
insertions is $\maxsumat{}{\subseqopen{A}{0}{n}}$ and is a candidate to
be the maximum subsequence sum of $A$.
To enumerate the remaining candidates,
the algorithm
performs additional operations on the list $L$, as shown at
lines~\ref{lin:forn2}--\ref{lin:forcheck}:
  for all $i \in \rangefromto{0}{n-2}$ in ascending order,
  (I)~remove the greatest member if its rank is $i$,
  (II)~accumulate \seqidx{A}{i}, and
  (III)~peek the greatest member.
The members of $L$ after the execution of the three operations for each $i$ are
determined by properties P1 and P2 with respect to the suffix sums $s_j =
\sum_{k \in \rangefromto{j}{i}} \seqidx{A}{k}$, $j \in \rangefromopento{0}{n}$.
They are in non-ascending order of value and in circular ascending order of
rank. As an example, note that the state of $L$ after the
execution of line~\ref{lin:push} at iteration $i = 0$ is
\begin{equation}
\left[
{
\sum_{k \in \rangefromto{4}{0}} \seqidx{A}{k} = 27,
\sum_{k \in \rangefromto{7}{0}} \seqidx{A}{k} = 24,
\sum_{k \in \rangefromto{14}{0}} \seqidx{A}{k} = 20,
\sum_{k \in \rangefromto{15}{0}} \seqidx{A}{k} = 12
}
\right],
\label{eq:exqueue}
\end{equation}
and that the greatest $L$'s member in this state is actually the
maximum sum of a nonempty subsequence ending at 0.

\begin{algorithm}[htbp]
\caption{\protect\CALL{maxSum}{A}: Computation of the maximum
subsequence sum of $A$ with a special queue.}
\begin{algorithmic}[1]
\State $MAX, i^*, j^* \gets 0$
\State $L \gets \emptyset$
\For{$i \gets 0, \ldots, n-1$}
	\State \CALL{accumulate}{L,A[i]} \label{lin:firstacc}
	\State \CALL{append}{L,A[i]} \label{lin:firstpush}
\EndFor
\State \CALL{checkForNewMax}{n-1}
\For{$i \gets 0, \ldots, n-2$} \label{lin:forn2}
	\If {$\CALL{peekRankOfMaxValue}{L} = i$}
		\State \CALL{remove}{L}
	\EndIf
	\State \CALL{accumulate}{L,A[i]} \label{lin:push}
	\State \CALL{checkForNewMax}{i} \label{lin:forcheck}
\EndFor
\Statex
\Procedure{checkForNewMax}{$i$}
	\If {$\CALL{peekMaxValue}{L} > MAX$}
		\State $MAX \gets \CALL{peekMaxValue}{L}$
		\State $i^* \gets \CALL{peekRankOfMaxValue}{L}$
		\State $j^* \gets i+1$
	\EndIf
\EndProcedure
\end{algorithmic}
\label{alg:maxsums}
\end{algorithm}

The implementation of the operations \CALL{remove}{}, \CALL{append}{},
and \CALL{peekRankOfMaxValue}{} as constant time functions is straightforward. To
devise an efficient implementation of \CALL{peekMaxValue}{}, the values of the
$L$'s members are stored indirectly, as follows. List $L$ is associated with the
value $D$ that equals the sum of all values appended with a
call to \CALL{append}{}.
Additionally, every $L$'s member $v$ is represented by the pair (its rank,
the value $D(v)$ that gives the value of $D$ just before the insertion of $v$).
The constant time execution of the \CALL{peekMaxValue}{} operation is trivial by simply returning $D - D(v)$,
for the first member $v$. The representation of the state in~\eqref{eq:exqueue}
is
\[
D=1, \ \
\left[
{
(-26,4),
(-23,7),
(-19,14),
(-11,15)
}
\right].
\]
The \CALL{accumulate}{} operation requires a linear time partial traversal of $L$ to preserve its nondecreasing order. However, this operation has amortized constant time complexity since every member is traversed at most once.

Once a maximum sum subsequence is characterized by $i^*$ and $j^*$, the
computation of $S_A$ in linear time is straightforward by means of algorithm
\CALL{MLS}{}. Therefore, we get

\begin{theorem}
The preprocessing algorithm computes the summary of $A$ in Tab.~\ref{tableCirc} in
$\Oof{n}$ time for any sequence $A$ of size $n$.
\end{theorem}

\subsection{Query-answering Step}

The following lemma establishes a property of the summary of $A$ that allows the
computation of the maximum subsequence sum of a circular shift of $A$ in the
case that this circular shift starts within a maximum sum subsequence of $A$.

\begin{lemma}
If $i \in \subseqopen{A}{i^*}{j^*}$, then
  \[
  \maxsumof{ \circsubseq{ A }{ i }{ i-1 } } = S_A[i].
  \]
\label{lem:mlssa}
\end{lemma}

\begin{figure}[htb]
\centering
\begin{tikzpicture}
\edef\arrayNumElem{21}
\EdefAsResultOf{\lastIdx}{int(\arrayNumElem-1)}
\drawEmptyArray

\edef\Kp {{0, 0, 0, 2, 2, 2, 1, 1, 1, 1, 1, 1, 1, 1, 1, 1, 2, 2, 0, 0, 0}}

\foreach \x/\r in {0/i,3/\hat\j,14/i^*,17/\hat\i,\lastIdx/i-1} {
                               	\EdefAsResultOf{\fillOfSubseq}{\IntervalColors[\Kp[\x]]}
                                  \drawSubseq{\x}{\x}{}{}
                                  \putNodeInfo{\x}{}{$\r$}
                                }

\putNodeInfo{6}{}{$j^*$}
\EdefAsResultOf{\fillOfSubseq}{\IntervalColors[5]}
\drawSubseq{6}{13}{nonpositive prefixes and suffixes}{}
\underbraceNodes{3} {16}  {\subseqopen{A} {\hat\j} {\hat\i}}
\end{tikzpicture}
\caption{Proof of Lemma~\ref{lem:mlssa}.}
\label{fig:proof}
\end{figure}
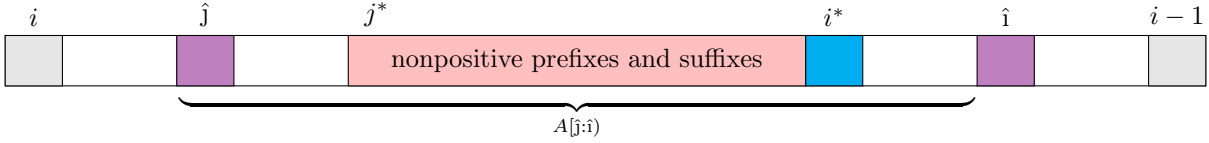

\begin{proof}
Since the sum of every suffix of $\subseqopen{A}{i^*}{j^*}$ is nonnegative, then
every maximum sum subsequence of $\subseqopen{A}{i}{j^*}$ is itself a suffix of
$\subseqopen{A}{i^*}{j^*}$, say $\subseqopen{A}{\hat \j}{j^*}$. Similarly, let
$\subseqopen{A}{i^*}{\hat \i}$ be a maximum sum subsequence of
$\subseqopen{A}{i^*}{i}$. It is worth noting that $\subseqopen{A}{\hat
\j}{\hat \i} = \subseqopen{A}{\hat
\j}{j^*}\subseqopen{A}{j^*}{i^*}\subseqopen{A}{i^*}{\hat \i}$. If
$\subseqopen{A}{q}{r}$ is a maximum sum subsequence of $\circsubseq{ A }{ i }{
i-1 }$, then $\rangefromopento{q}{r} \subseteq \rangefromopento{\hat \j}{\hat
\i}$ or $\sumof{\subseqopen{A}{q}{\hat \j}} = 0$ or $\sumof{\subseqopen{A}{\hat
\i}{r}} = 0$. It turns out that $\maxsumof{ \circsubseq{ A }{ i }{ i-1 } } =
\maxsumof{ \circsubseqopen{ A }{ \hat \j }{ \hat \i } }$, as depicted in
Fig.~\ref{fig:proof}. In addition, if $k \in \subseqopen{A}{j^*}{i^*}$, then
$\sumof{\subseqopen{A}{j^*}{k}} \leq 0$ and $\sumof{\subseqopen{A}{k}{i^*}} \leq 0$. Therefore, the lemma follows.
\end{proof}

We finally get the main result for the circular case based
on~\eqref{eq:circposx} and Lemma~\ref{lem:mlssa}.

\begin{theorem}
Given a summary as defined in Tab.~\ref{tableCirc},
  the query-answering algorithm computes \maxcircsumof{ \seqins[x]{A}{p} } in
  \Oof{1} time for any sequence $A$, $x \in \SetOfRealNumbers$, and $p \in
  \rangefromopento{0}{n}$ as
\[
\maxcircsumof{\seqins{A}{p}} = \left\{ \begin{array}{ll}
\max\set{S[p], x + MAX}, &
\text{if } p \in \rangefromopento{i^*}{j^*} \\
\max\set{ MAX, x + S[p] }, &
\text{otherwise.} \end{array} \right.
\]
\end{theorem}
%

%% file: FigGenApproachCircBeforeIns.tex
\edef\arrayNumElem{16}
\EdefAsResultOf{\lastIdx}{int(\arrayNumElem-1)}

\EdefAsResultOf{\baseAngle}{(360/\arrayNumElem)}
\EdefAsResultOf{\halfBaseAngle}{(\baseAngle/2)}
\EdefAsResultOf{\circumference}{(\arrayNumElem*\elemXSide)}
\EdefAsResultOf{\intRadius}{(\circumference/(2*3.1415))}
\EdefAsResultOf{\extRadius}{(\intRadius+\elemYSide)}
\EdefAsResultOf{\middleRadius}{(\intRadius+(\elemYSide/2))}

\foreach \x in {0,...,\lastIdx} {
				\EdefAsResultOf{\angle}{-(\x*\baseAngle)-180-(\baseAngle)}
                \EdefAsResultOf{\number}{int(\A[\x])}
                \EdefAsResultOf{\fillOfSubseq}{\IntervalColors[\Kc[\x]]}

\draw[draw=\drawOfSubseq,fill=\fillOfSubseq] (\angle:\extRadius) --
(\angle+\baseAngle:\extRadius) -- (\angle+\baseAngle:\intRadius) --
(\angle:\intRadius) -- (\angle:\extRadius);
\draw[draw=\drawOfSubseq] (\angle+\halfBaseAngle:\middleRadius) node{\number};
}

\EdefAsResultOf{\outRadius}{\extRadius+(\elemYSide/2)}

\foreach \x in {0,...,\lastIdx} {
\EdefAsResultOf{\angle}{-(\x*\baseAngle)-180-(\baseAngle)}
\node at (\angle+\halfBaseAngle:\outRadius)
{\ifnum \x = 4
	$i^*$
\else \ifnum \x = 1
	$j^*$
\else
 	\x
\fi
\fi}; }

%% file: FigGenApproachCircCompBeforeIns.tex
\edef\arrayNumElem{16}
\EdefAsResultOf{\lastIdx}{int(\arrayNumElem-1)}

\edef\Kcc {{0, 0, 0, 0, 0, 0, 0, 0, 0, 0, 0, 0, 1, 1, 1, 0}}

\EdefAsResultOf{\baseAngle}{(360/\arrayNumElem)}
\EdefAsResultOf{\halfBaseAngle}{(\baseAngle/2)}
\EdefAsResultOf{\circumference}{(\arrayNumElem*\elemXSide)}
\EdefAsResultOf{\intRadius}{(\circumference/(2*3.1415))}
\EdefAsResultOf{\extRadius}{(\intRadius+\elemYSide)}
\EdefAsResultOf{\middleRadius}{(\intRadius+(\elemYSide/2))}

\foreach \x in {0,...,\lastIdx} {
\EdefAsResultOf{\angle}{-(\x*\baseAngle)-180-(\baseAngle)}
                                  \EdefAsResultOf{\number}{int(-\A[\arrayNumElem-1-\x])}
                                  \EdefAsResultOf{\fillOfSubseq}{\IntervalColors[\Kcc[\x]]}

\draw[draw=\drawOfSubseq,fill=\fillOfSubseq] (\angle:\extRadius) --
(\angle+\baseAngle:\extRadius) -- (\angle+\baseAngle:\intRadius) --
(\angle:\intRadius) -- (\angle:\extRadius);
\draw[draw=\drawOfSubseq] (\angle+\halfBaseAngle:\middleRadius) node{\number};
}

\EdefAsResultOf{\outRadius}{\extRadius+(\elemYSide/2)}

%
\foreach \x in {0,...,\lastIdx} {
\EdefAsResultOf{\angle}{-(\x*\baseAngle)-180-(\baseAngle)}
\node at (\angle+\halfBaseAngle:\outRadius)
{\ifnum \x = 12
	$n-i^*$
\else \ifnum \x = 15
	$\!\!\!\! n-j^*$
\else
 	\x
\fi
\fi}; }

%% file: FigGenApproachCircAfterIns.tex
\edef\arrayNumElem{17}
\EdefAsResultOf{\lastIdx}{int(\arrayNumElem-1)}

\edef\Apc{{2, -7, 4, 28, -25, 12, -1, -8, 14, 1, -6, -3, 5, 11, -18, 8, 10}}
\edef\Kpc {{1, 1, 1, 2, 0, 1, 1, 1, 1, 1, 1, 1, 1, 1, 1, 1, 1}}

\EdefAsResultOf{\baseAngle}{(360/\arrayNumElem)}
\EdefAsResultOf{\halfBaseAngle}{(\baseAngle/2)}
\EdefAsResultOf{\circumference}{(\arrayNumElem*\elemXSide)}
\EdefAsResultOf{\intRadius}{(\circumference/(2*3.1415))}
\EdefAsResultOf{\extRadius}{(\intRadius+\elemYSide)}
\EdefAsResultOf{\middleRadius}{(\intRadius+(\elemYSide/2))}

\foreach \x in {0,...,\lastIdx} {
\EdefAsResultOf{\angle}{-(\x*\baseAngle)-180-(\baseAngle)}
                                  \EdefAsResultOf{\number}{int(\Apc[\x])}
                                  \EdefAsResultOf{\fillOfSubseq}{\IntervalColors[\Kpc[\x]]}

\draw[draw=\drawOfSubseq,fill=\fillOfSubseq] (\angle:\extRadius) --
(\angle+\baseAngle:\extRadius) -- (\angle+\baseAngle:\intRadius) --
(\angle:\intRadius) -- (\angle:\extRadius);
\draw[draw=\drawOfSubseq] (\angle+\halfBaseAngle:\middleRadius) node{\number};
}

\EdefAsResultOf{\outRadius}{\extRadius+(\elemYSide/2)}

\foreach \x in {0,...,\lastIdx} {
\EdefAsResultOf{\angle}{-(\x*\baseAngle)-180-(\baseAngle)}
                                  \EdefAsResultOf{\mcs}{int(\x)}

\draw[draw=\drawOfSubseq] (\angle+\halfBaseAngle:\outRadius) node{\mcs};
}

%% file: SecUnitConclusion.tex
\section{Concluding Remarks} \label{SecUnitConclusion}%

In this paper we have considered the problem of,
  for a fixed sequence $A$ of $n$ real numbers,
  answering queries which ask the value
  of \maxsumof{\seqins[x]{A}{p}} or \maxcircsumof{\seqins[x]{A}{p}}
  for given $x \in \SetOfRealNumbers$ and $p \in \rangefromto{0}{n}$.
We showed that,
  after an \Oof{n} time preprocessing step has been carried out on $A$,
  both kinds of queries can be answered in constant worst-case time.
This is both an optimal solution to the problem
  and a considerable improvement over the naive strategy
  of answering such queries by means of Kadane's algorithm
  (or a variation of it, in the circular case),
  which takes \Thetaof{n} time per query.
This problem has applications in the context of finding heuristic solutions to
an NP-hard problem of buffer minimization in wireless mesh networks.
Given the generality of these kinds of queries
  and the multiplicity of applications of the maximal sum subsequence concept,
  we would not be surprised to see other applications of our algorithms in the future.
An interesting related problem is then that of,
  given an $m \times n$ matrix $A$,
  inserting $k$ size-$m$ columns $C_0, \ldots, C_{k-1}$
  successively and cumulatively into $A$ in order to minimize
  $\costofmatrix{A'} = \sum_{i=0}^{m-1} \maxcircsumof{ \seq{
  \matrixelem{A'}{i}{0}, \ldots, \matrixelem{A'}{i}{n-1} } }$ for every matrix
  $A'$ resulting of each insertion.
By using the algorithms presented in this paper
  to insert one column at a time,
  one can carry out $k$ successive insertions in
  \( \Oof{ m(n+1) + m(n+2) + \ldots + m(n+k) }
     = \Oof{ k(mn + mk) }
    \)
  time.
However,
  since the input to the problem has size $\Oof{mn + mk}$,
  we leave it as an open problem whether substantially more efficient algorithms exist.

%% file: main_arxiv.bbl
\begin{thebibliography}{10}

\bibitem{BibAkyildizWangWang2005}
I.~F. Akyildiz, X.~Wang, and W.~Wang.
\newblock Wireless mesh networks: a survey.
\newblock {\em Computer Networks}, 47(4):445--487, 2005.

\bibitem{BibKlasingMoralesPerennes2008}
R.~Klasing, N.~Morales, and S.~P{\'e}rennes.
\newblock On the complexity of bandwidth allocation in radio networks.
\newblock {\em Theoretical Computer Science}, 406(3):225--239, 2008.

\bibitem{BibVieiraRezendeBarbosaFdida2012}
F.~R.~J. Vieira, J.~F. de~Rezende, V.~C. Barbosa, and S.~Fdida.
\newblock Scheduling links for heavy traffic on interfering routes in wireless
  mesh networks.
\newblock {\em Computer Networks}, 56(5):1584--1598, 2012.

\bibitem{BibLeModianoShroff2012}
L.~B. Le, E.~Modiano, and N.~B. Shroff.
\newblock Optimal control of wireless networks with finite buffers.
\newblock {\em IEEE/ACM Transactions on Networking}, 20(4):1316--1329, 2012.

\bibitem{BibBentley1984}
J.~Bentley.
\newblock Programming pearls: algorithm design techniques.
\newblock {\em Communications of the ACM}, 27(9):865--873, 1984.

\bibitem{BibAnPeursumLiuVenkatesh2009}
S.~An, P.~Peursum, W.~Liu, and S.~Venkatesh.
\newblock Efficient algorithms for subwindow search in object detection and
  localization.
\newblock In {\em Proc. of IEEE Conference on Computer Vision and Pattern
  Recognition}, pages 264--271, 2009.

\bibitem{BibFukudaMorimotoMorishitaTokuyama2001}
T.~Fukuda, Y.~Morimoto, S.~Morishita, and T.~Tokuyama.
\newblock Data mining with optimized two-dimensional association rules.
\newblock {\em ACM Transactions on Database Systems}, 26(2):179--213, 2001.

\bibitem{BibRuzzoTompa1999}
W.~L. Ruzzo and M.~Tompa.
\newblock A linear time algorithm for finding all maximal scoring subsequences.
\newblock In {\em Proc. of the Seventh International Conference on Intelligent
  Systems for Molecular Biology}, pages 234--241, 1999.

\bibitem{BibLinJiangChao2002}
Y.-L. Lin, T.~Jiang, and K.-M. Chao.
\newblock Efficient algorithms for locating the length-constrained heaviest
  segments with applications to biomolecular sequence analysis.
\newblock {\em Journal of Computer and System Sciences}, 65(3):570--586, 2002.

\bibitem{BibCsuros2004}
M.~Csűrös.
\newblock Maximum-scoring segment sets.
\newblock {\em IEEE/ACM Transactions on Computational Biology and
  Bioinformatics}, 1(4):139--150, 2004.

\bibitem{BibFukudaTakaoka2007}
K.~Fukuda and T.~Takaoka.
\newblock Analysis of air pollution ($\mathrm{PM}_{10}$) and respiratory
  morbidity rate using {K}-maximum sub-array (2-{D}) algorithm.
\newblock In {\em Proc. of the ACM Symposium on Applied Computing}, pages
  153--157, 2007.

\bibitem{BibThaherTakaoka2010}
M.~Thaher and T.~Takaoka.
\newblock An efficient algorithm for the k maximum convex sums.
\newblock {\em Procedia Computer Science}, 1(1):1475--1483, 2010.

\bibitem{BibLu2010}
W.~Lu.
\newblock Improved {SWOT} approach for conducting strategic planning in the
  construction industry.
\newblock {\em Journal of Construction Engineering and Management},
  136(12):1317--1328, 2010.

\bibitem{BibGries1982}
D.~Gries.
\newblock A note on a standard strategy for developing loop invariants and
  loops.
\newblock {\em Science of Computer Programming}, 2(3):207--214, 1982.

\bibitem{BibBentley1984November}
J.~Bentley.
\newblock Programming pearls: perspective on performance.
\newblock {\em Communications of the ACM}, 27(11):1087--1092, 1984.

\bibitem{BibThaherTakaoka2012}
M.~Thaher and T.~Takaoka.
\newblock Improved algorithms for the k overlapping maximum convex sum problem.
\newblock {\em Procedia Computer Science}, 9:754--763, 2012.

\bibitem{BibChenChao2007}
K.-Y. Chen and K.-M. Chao.
\newblock On the range maximum-sum segment query problem.
\newblock {\em Discrete Applied Mathematics}, 155(16):2043--2052, 2007.

\bibitem{BibDaiSu2006}
H.-K. Dai and H.-C. Su.
\newblock A parallel algorithm for finding all successive minimal maximum
  subsequences.
\newblock In J.~R. Correa, A.~Hevia, and M.~Kiwi, editors, {\em Proc. of the
  Latin American Theoretical Informatics}, volume 3887 of {\em Lecture Notes in
  Computer Science}, pages 337--348, 2006.

\bibitem{BibAlvesCaceresSong2006}
C.~E.~R. Alves, E.~N. C\'{a}ceres, and S.~W. Song.
\newblock A {BSP/CGM} algorithm for finding all maximal contiguous subsequences
  of a sequence of numbers.
\newblock In W.~E. Nagel, W.~V. Walter, and W.~Lehner, editors, {\em Proc. of
  the 12th International Euro-Par Conference}, volume 4128 of {\em Lecture
  Notes in Computer Science}, pages 831--840, 2006.

\bibitem{BibBrodalJorgensen2007}
G.~S. Brodal and A.~G. Jørgensen.
\newblock A linear time algorithm for the $k$ maximal sums problem.
\newblock In L.~Kučera and A.~Kučera, editors, {\em Proc. of 32nd
  International Symposium}, volume 4708 of {\em Lecture Notes in Computer
  Science}, pages 442--453, 2007.

\bibitem{BibHuang1994}
X.~Huang.
\newblock An algorithm for identifying regions of a {DNA} sequence that satisfy
  a content requirement.
\newblock {\em Computer Applications in the Biosciences (CABIOS)},
  10(3):219--225, 1994.

\bibitem{BibMu2008}
S.-C. Mu.
\newblock Maximum segment sum is back: Deriving algorithms for two segment
  problems with bounded lengths.
\newblock In {\em Proc. of the ACM SIGPLAN Symposium on Partial Evaluation and
  Semantics-based Program Manipulation}, pages 31--39, 2008.

\bibitem{BibChengLiuChao2009}
C.-H. Cheng, H.-F. Liu, and K.-M. Chao.
\newblock Optimal algorithms for the average-constrained maximum-sum segment
  problem.
\newblock {\em Information Processing Letters}, 109(3):171--174, 2009.

\bibitem{BibBrodalJorgensen2008}
G.~S. Brodal and A.~G. Jørgensen.
\newblock Selecting sums in arrays.
\newblock In S.-H. Hong, H.~Nagamochi, and T.~Fukunaga, editors, {\em
  Algorithms and Computation}, volume 5369 of {\em Lecture Notes in Computer
  Science}, pages 100--111. 2008.

\bibitem{BibLiHui1992}
T.~Li-Hui.
\newblock Sequencing to minimize the maximum renewal cumulative cost.
\newblock {\em Operations Research Letters}, 12(2):117--124, 1992.

\bibitem{BibCorreaFariasSouza2013}
R.~C. Corr{\^e}a, P.~M.~S. Farias, and C.~P. de~Souza.
\newblock Insertion and sorting in a sequence of numbers minimizing the maximum
  sum of a contiguous subsequence.
\newblock {\em Journal of Discrete Algorithms}, 21:1--10, 2013.

\bibitem{Jeuring1991}
J.~Jeuring.
\newblock Incremental algorithms on list.
\newblock In {\em Proc. SION Computing Science in the Netherlands,}, pages
  315--335, 1991.

\end{thebibliography}
